\documentclass[acmsmall]{acmart}

\AtBeginDocument{%
  }

\setcopyright{acmlicensed}
\copyrightyear{2025}
\acmYear{2025}
\acmDOI{XXXXXXX.XXXXXXX}

\acmConference[To appear]{ACM SIGCHI Conference on Computer-Supported Cooperative Work \& Social Computing}{}{}

\acmISBN{}

\usepackage[flushleft]{threeparttable}
\appto\TPTdoTablenotes{\scriptsize}
\usepackage{graphicx}
\usepackage{multirow}
\usepackage{booktabs}
\usepackage{multirow,longtable}
\usepackage{enumitem}

\begin{document}

\title[Turn-Taking Behavior in Open-Ended Group Activities in Virtual Reality]{Predicting and Understanding Turn-Taking Behavior in Open-Ended Group Activities in Virtual Reality}


\author{Portia Wang}
\affiliation{%
 \institution{Stanford University}
 \city{Stanford}
 \state{California}
 \country{USA}}
\email{portiaw@stanford.edu}

\author{Eugy Han}
\affiliation{%
 \institution{Stanford University}
 \city{Stanford}
 \state{California}
 \country{USA}}

\author{Anna C.M. Queiroz}
\affiliation{%
 \institution{Stanford University}
 \city{Stanford}
 \state{California}
 \country{USA}}

\author{Cyan DeVeaux}
\affiliation{%
 \institution{Stanford University}
 \city{Stanford}
 \state{California}
 \country{USA}}

\author{Jeremy N. Bailenson}
\affiliation{%
 \institution{Stanford University}
 \city{Stanford}
 \state{California}
 \country{USA}}

\renewcommand{\shortauthors}{Wang et al.}

\begin{abstract}

In networked virtual reality (VR), user behaviors, individual differences, and group dynamics can serve as important signals into future speech behaviors, such as who the next speaker will be and the timing of turn-taking behaviors. The ability to predict and understand these behaviors offers opportunities to provide adaptive and personalized assistance, for example helping users with varying sensory abilities navigate complex social scenes and instantiating virtual moderators with natural behaviors. In this work, we predict turn-taking behaviors using features extracted based on social dynamics literature. We discuss results from a large-scale VR classroom dataset consisting of 77 sessions and 1660 minutes of small-group social interactions collected over four weeks. In our evaluation, gradient boosting classifiers achieved the best performance, with accuracies of 0.71--0.78 AUC (area under the ROC curve) across three tasks concerning the ``what’’, ``who’’, and ``when’’ of turn-taking behaviors. In interpreting these models, we found that group size, listener personality, speech-related behavior (e.g., time elapsed since the listener’s last speech event), group visual attention (e.g., the group's head orientation towards the speaker), and the listener and previous speaker’s head pitch, head y-axis position, and left hand y-axis position more saliently influenced predictions. Results suggested that these features remain reliable indicators in novel social VR settings, as prediction performance is robust over time and with groups and activities not used in the training dataset. We discuss theoretical and practical implications of the work.

\end{abstract}




\maketitle

\section{Introduction}

Effective and organized social interactions require its social actors to accurately infer speaking intention and take turns accordingly. In most platforms for facilitating computer mediated-communication (CMC), users retain a high level of autonomy over when and how they speak, utilizing features such as the muting and unmuting button during videoconferencing. However, along with the benefits of user control comes the responsibility for individuals to correctly infer social group dynamics. Such inferences can be difficult to navigate as users utilize digital platforms such as video conferencing and social virtual reality (VR), where the level of immersion and representation of users (e.g., nonverbal behavior, avatar appearance) can differ from one another \cite{Steinicke2020, Abdullah2021}. In immersive social interactions, differences in avatar representation and characteristics of social VR platforms can yield adverse outcomes such as unwanted overlap in speech \cite{BarredaAngeles2023, Williamson2021}, lack of control for users with physical disabilities \cite{Maloney2020}, misperceived and absence of nonverbal cues \cite{Maloney2020, Moustafa2018}, and difficulty gauging the end of speaking turns \cite{Moustafa2018, Williamson2021}. For users with visual and audio disabilities, it can also be challenging for them to engage in seamless interactions based solely off of perceptual and auditory feedback \cite{Kushalnagar2020, Anderson2021}. Paradoxically, as immersive communication evolves to be hardware-agnostic, with users interacting across traditional screen devices and mixed-reality headsets \cite{Qiu2023}, the mismatch in social cues delivered and perceived by users across different devices present further complications for users to navigate socially.

Being able to predict user intentions for actions such as turn taking can alleviate these challenges through real-time interventions and training. Social VR platforms can implement personalized assistance for users who are struggling to interpret nonverbal cues under technological constraints (e.g., joining large meetings though audio feed) or are different in physical abilities (e.g., auditory and visual abilities). This assistance can also benefit populations with medical conditions such as autism and ADHD that inhibit their ability to accurately interpret nonverbal behaviors \cite{Freeman2002, Hall1999}. Relatedly, understanding \emph{what} behavior precedes turn taking can be useful in training practitioners such as educators and healthcare workers to better navigating social settings. For example, by training early-career practitioners to recognize nonverbal cues and individually characteristics influencing speaking intentions, they can be better equipped to navigate social situations.

In more structured settings such as classrooms and focus groups, knowing who will likely speak next and what nonverbal cues precede speech can improve social dynamics. For one, practitioners who need to monitor multiple virtual sessions simultaneously, inevitably missing nonverbal and verbal cues, can leverage turn-taking predictions to remain informed of the social dynamics and better facilitate group conversations. These practitioners can also instantiate virtual agents to moderate social VR scenes with natural nonverbal behaviors that do not cut off current speakers and more broadly design mechanisms for smoothly guiding the conversations. Failure to accurately infer social dynamics can yield unwanted interruptions and lower the perceived ease of use of these intelligent systems \cite{Maier2022CA}. As such, we argue that it is important to study whether we can \emph{robustly} predict virtual turn-taking behaviors and how nonverbal and verbal behaviors as well as individual and group characteristics influence turn-taking predictions. 

In this work, we investigated turn-taking behaviors in VR and leveraged the medium’s fine-grain tracking data on motion and verbal behaviors. Using the VR motion data, researchers have implemented pipelines for deriving features such as head movement \cite{Miller2023VR} and gaze \cite{Abdullah2021}, both of which have been previously useful for predicting speech behaviors in face-to-face social interactions \cite{Ishii2017Head, Ishii2016Gaze}. As such, our first research question considers whether it is possible to predict turn taking.

\begin{itemize}

\item Research Question 1 (\textbf{RQ1}): Can we predict turn-taking behaviors in VR open-ended group activities from features extracted from motion tracking data, speech-related behavior, and individual and group differences?

\end{itemize}

Building on this, we leveraged VR to place individuals into varying virtual spaces and studied social dynamics across groups, social activities, and spatial contexts over time. Studying diverse social interactions at scale is important as building models that are robust to unseen groups, activities, and time suggests higher practicality and reliability. Concretely, our second research question concerns generalizability.

\begin{itemize}
\item Research Question 2 (\textbf{RQ2}): How is turn-taking behavior prediction performance affected when evaluated on groups, activities, and time not seen in training?
\end{itemize}

Finally, we interpreted the predictive models. Analyzing how features predict turn taking can deepen our understanding of virtual social dynamics, allowing future researchers to benchmark against face-to-face interactions. This understanding can also help practitioners build more efficient models with curated features and compose social groups exhibiting certain turn-taking behaviors.

\begin{itemize}

\item Research Question 3 (\textbf{RQ3}): How are the extracted features related to turn-taking predictions and performance? Specifically, what features are strongly related to performance, and how are they associated with turn-taking predictions?

\end{itemize}

To answer these questions, we built models predicting speech dynamics, focusing on three tasks that capture the ``what’’, ``who’’, and ``when’’ of turn-taking behaviors. We formulated features based on tracking data and individual and group characteristics based on social dynamics literature. We evaluated performance across 77 sessions and 1660 minutes of open-ended VR group discussions spanning four weeks, each with three to four university students. Notably, we found that gradient boosting classifiers achieved the best accuracies, predicting turn transitions and identifying the new speaker with 0.75--0.78 AUC (i.e., area under the ROC curve), and differentiating between moments immediately preceding turn transitions and those sampled at prior moments with accuracies of 0.71--0.72 AUC. These results demonstrated that we were able to predict turn-taking behaviors with accuracies considerably higher than prediction by chance (i.e., 0.50 AUC) and highlighted the potential non-linear relationships and interactions between extracted features.

Additional feature analyses revealed the importance of listener personality, group size, speech-related behavior (e.g., preceding speaker sequence), group visual attention, head pitch and y-axis position, and left hand y-axis position. Our models showed comparable results when evaluated on time, activities, and groups not seen during training, which demonstrated the reliability of these turn-taking indicators in novel social settings. From these results, we highlight their theoretical relevance and outline practical implications for how practitioners can leverage behavioral predictions to facilitate effective VR social interactions.

In summary, we make the following contributions. First, we formulated features based on social dynamics literature and leveraged both the VR tracking data and individual and group characteristics. Using a large-scale VR dataset with 1660 minutes of open-ended activities collected over four weeks, we then demonstrated the feasibility and robustness of predicting VR turn-taking behaviors. Finally, through interpreting how features influence model performance and prediction probabilities, we present theoretical and practical insights and highlight the potential of our work for intervention, support, and training.

\section{Related Work} \label{Sec: Related work}


\subsection{Nonverbal Behavior in Social Interactions} \label{Sec: Nonverbal Behavior in Social Interactions}

Nonverbal behavior offers insights into social dynamics. Gaze, for example, is related to action patterns \cite{Kendon1969} and attention \cite{Kutt2020, Semsar2020}, and provides important signals to understanding dyadic collaborations \cite{Andrist2015, Jing2022}, user intention and coordination \cite{Amati2018}, learning outcomes \cite{Pi2020}, conversational attention \cite{Vertegaal2001}, and social engagement \cite{Nakano2010}. Gesture and body orientation can influence the sense of feeling addressed by conversation partners \cite{Nagels2015}, while proxemics, body orientation, and gaze, can be effective in predicting one’s intention of joining social groups \cite{Bonsch2020}.

It is then unsurprising that nonverbal communication makes up a key component of virtual interactions, for which social VR offers additional benefits compared to video conferencing. By tracking the users' eyes, heads, and hands, VR systems render avatars that accurately convey nonverbal cues such as interpersonal distance and gaze, fostering social interactions that are perceived positively and comparable to face-to-face interactions \cite{Maloney2020, Smith2018}. Communication through nonverbal behaviors such as bodily and facial gestures induce more positive dyadic interactions in VR, with these cues also being predictive of interpersonal attraction \cite{Oh2020}. Using motion data, VR social interactions can also be augmented, for example through increasing mimicry \cite{Tarr2018, Raffard2018, Bailenson2005} and transforming gaze and interpersonal distance \cite{Roth2018, Roth2019, Wang2024BG,Wang2023BG}. With VR placing users in controllable virtual environments, it is also possible to examine nonverbal behavior longitudinally \cite{Miller2023VR, Miller2023Identification}. Works highlighted how nonverbal behavior changes over time – users looked at others more and their interpersonal distance increased \cite{Miller2023VR}, and user identifiability lowered with greater temporal delay between training and testing sessions \cite{Miller2023Identification}.

VR tracking data also enables fine-grain analyses on synchrony \cite{Sun2019, Miller2021Synchrony}, self-efficacy and learning \cite{Queiroz2023}, physiological responses \cite{Luong2022}, classroom discourse \cite{Stark2024}, interpersonal distance \cite{Miller2023VR, DeVeaux2024}, context \cite{Han2024Spatial}, design behaviors \cite{Wang2024DS}, and user identification \cite{Miller2020, Nair2023, Miller2022, Miller2023Identification, Moore2021}. Particularly relevant to our work is the research of DeVeaux et al. \cite{DeVeaux2024}. The authors extracted linguistic patterns from transcripts and notably found a positive correlation between the use of impersonal pronouns and the median interpersonal distance. This research highlighted the importance of considering language use to study VR affordances and the potential of leveraging nonverbal behavior to uncover psychological nuances that surveys fail to capture. Key distinctions between their work and ours are their focus on the nonverbal behavior of interpersonal distance, linguistic styles, and session-level analyses. In contrast, we studied turn-taking behaviors and focused on predicting them at moments within sessions. Beyond interpersonal distance, we also included other nonverbal features (e.g., egocentric motion). One other distinction lies in the VR activities analyzed: while DeVeaux et al. \cite{DeVeaux2024} studied instructor-led discussions collected during a university course in Fall 2021, we focused on topen-ended activities recorded a year later through the same course.

We contribute to past works by predict turn-taking behaviors through VR motion data. Findings on changes in nonverbal behaviors over time \cite{Miller2023VR, Miller2023Identification} emphasized the need to study social behaviors through a longitudinal lens and across multiple sessions. These insights motivated our examination of turn-taking predictions across unseen weeks, groups, and activities.

\subsection{Modeling Human Behavior and Individual-  and Group- Level Differences using Tracking Data} \label{Sec: Modeling using Tracking Data}

Besides using tracking data to understand human behaviors, monitoring user motion, often unobtrusively, can further enable interventions and assessments. For example, as smartphone data such as touch and typing behaviors \cite{Wampfler2022, Kovavcevic2023} and audio, textual, and video data \cite{Kampman2018} are predictive of personality traits and affective states, one can construct personalized experiences based on the user’s current behaviors. Another example involves monitoring gaze to assess cognitive load, and using this information to facilitate adaptations of mixed reality interfaces \cite{Lindlbauer2019}. Recently, researchers also built models for predicting team viability using online team text conversations and demonstrated the potential of using automated features for assessments and intervention \cite{Cao2021}. Others proposed using individual and social behaviors to model purchase decisions for e-commerce recommendations \cite{Xu2019} and job burnouts for early-stage interventions \cite{Wu2021}.

Similarly, we see opportunities in predicting VR speech behaviors for intervention and assistance. Fortunately, though humans can struggle to infer speaker behavior \cite{Heeman2015}, research that leverages tracking data for speech behavior predictions has shown greater promise. Works looked to speech sequences \cite{Parker1988}, mouth and head motion \cite{Ishii2016Mouth, Ishii2017Head}, respiration \cite{Ishii2016Respiration}, gaze behavior \cite{Ishii2016Gaze, Lee2023}, and more broadly user motion \cite{Chen2024} as predictors when modeling speech behavior. Notably, these works have found that markov models were effective in predicting the next speaker based on the two preceding speakers \cite{Parker1988} and that support vector machines leveraging features on gaze transition patterns are predictive of turn-changing characteristics \cite{Ishii2016Gaze}.

Unlike the social scenes examined in previous studies, where users completed a single or series of similar tasks \cite{Chen2024,Ishii2016Gaze} and either stood in fixed positions \cite{Lee2023} or remained seated \cite{Chen2024, Ishii2016Gaze}, VR social interactions are diverse in activities and typically involve less constraints on movement in the physical and virtual spaces. This raises the question of whether turn-taking predictions are possible when there are greater variations in activities and nonverbal cues. Failure to make robust predictions in these scenarios limits the prospects of intervention and assistance in VR. To address this, we investigated VR turn-taking behaviors from four weeks of open-ended activities with little restriction on virtual motion. We contribute by also interpreting how features are related to turn-taking predictions, highlighting their theoretical and practical significance.

\subsection{Social Interactions across Individuals and Groups}

Tracking data's ability to predict individual differences such as personalities and affective states \cite{Wampfler2022, Kovavcevic2023, Kampman2018, Ivanov2011} suggests that individual differences can be used to predict user behaviors. Individual differences such as gender, age, and personalities are related to how people take up physical and virtual spaces \cite{Friebel2021, Iachini2016GenderAge, Nassiri2010, Iachini2014Gender}, while personality, public speaking anxiety, and immersive tendencies of individuals are predictive of user experience and perceived quality of interactive systems\cite{Chollet2018}. 

Personality traits, in particular, are related to speech behaviors. For example, conscientiousness and extraversion are related to basic speech features such as pitch estimate \cite{Ivanov2011}. Extraversion and neuroticism are also associated with different speech patterns, notably with introverts exhibiting longer silences between utterances \cite{Ramsay1968}. Introverts and extroverts were also found to speak at different levels of language abstraction \cite{Beukeboom2013} and differ in reaction time when verbally responding to prompts and questionnaires \cite{Park2020, Lee2021}. Broadly speaking, extraversion, agreeableness, and neuroticism also correlate with features related to speaking turns, speaking length, and average speaking turn duration \cite{Aran2013, Lepri2010}.

Broadly speaking, social interactions can vary depending on the characteristics of social groups. For example, the association between trait dominance---measured as expressed control and prosocial interpersonal power and influence---and speaking time is influenced by the composition of social groups, namely whether the group was composed randomly or of extremes (i.e., pairing together the most and least dominant individuals) \cite{Mast2002}. Comparing conversations between dyad and triads, the presence of an additional listener led to individuals speaking louder in situations with high noise levels, and listeners orienting their heads more optimally (i.e., rotating their heads to approximately 30 degrees from the speaker) \cite{Hadley2021}.

We rely on past findings to identify features related to personality and group compositions for modeling turn taking. By studying how these features are related to social behaviors such as speaking intentions, we explore how individual and group characteristics can help build immersive tools for facilitating social interactions.

\begin{figure*}[t]
 \centering 
 \includegraphics[width=0.8\textwidth,keepaspectratio]{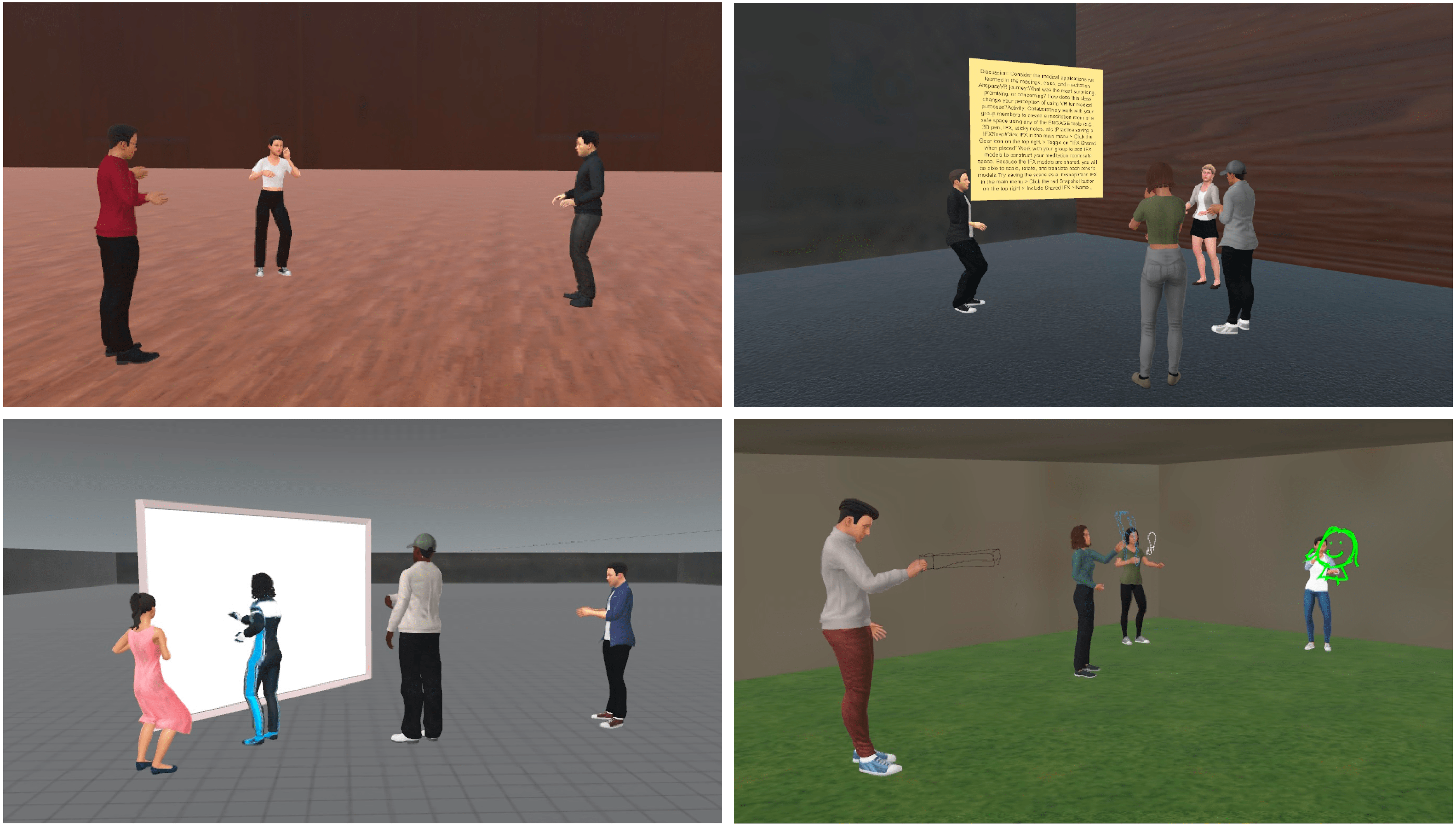}
 \caption{Screenshots of the Open-Ended Group Activities. Students donned VR gear in remote physical locations but joined together in different virtual environments for weekly group discussions and participated in activities related to topics on accessibility, avatars, medical, and education.}
 \label{fig: engage}
\end{figure*}

\section{Methods}
\subsection{Recordings of Open-ended Group Social Interactions in VR}

We studied the longitudinal VR classroom dataset collected by Han et al. \cite{Han2024}, and focused on the subset collected in Fall 2022 as part of a university course on VR. In it, 146 university students, out of whom 117 consented, engaged in weekly discussions and responded to open-ended design prompts in groups of two to four for four weeks. Given our interest in turn taking, we only analyzed VR sessions with all consenting participants\footnote{A research personnel not part of the teaching staff randomly assigned students into groups and maximized the number of groups with all consenting students. This ensured that students received comparable learning experience as the teaching staff was blinded to the consenting status, while also allowed researchers to examine fully-consenting group interactions.}. The topics of discussion and activities changed weekly, with themes such as accessibility and education. We detail session activities in Appendix \ref{dataset details}. Each week, students gathered in the same groups and used the social VR platform ENGAGE while being physically remote in their own private spaces. Each student used a Meta Quest 2 headset and two hand-held controllers to partake in the sessions and was allowed to move around virtually using smooth translation and teleportation. No teaching staff was present in these sessions, so student groups recorded the discussion sections using the platform’s recording feature. Figure \ref{fig: engage} shows screenshots from the recorded group activities.

The authors varied the virtual environments (i.e., ceiling height, amount of visible space) to examine how context influences attitude, nonverbal behavior, and design behavior \cite{Han2024Spatial, Wang2024DS}. This variation in context allows us to model interactions occurring in diverse virtual spaces, setting it apart from literature that examined speech behavior in the same physical or virtual environment \cite{Chen2024, Lee2023}. This particular dataset was also well-suited for our purpose given its open-ended activities, which fostered more natural interactions compared to the instructor-led sections in the rest of the dataset. While previous research have examined language use using the data collected in Fall 2021 \cite{DeVeaux2024, DeVeaux2023}, we are the first to use the Fall 2022 data to examine turn-taking behaviors of open-ended group activities.

Given our interest in group interactions, we filtered out moments and sessions with two students. For predicting turn-taking behaviors, looking at groups larger than two also makes the problem nontrivial. Our final dataset consisted of 77 VR sessions collected from 26 groups and 100 unique students. Out of the 77 sessions, 35 were three-person discussions, and 42 were of four people. During data collection, some participants missed discussions, leading to a high number of three-person discussions. Some sessions were dropped due to technical difficulties such as software updates and incomplete recordings. The dataset aggregated a total of 1660.60 minutes of tracking data, with each session taking on average 21.57 minutes (SD=6.44).

The recording files collected several forms of user and session data, namely that of the scene, user motion, and audio. Since we wanted to extrapolate turn-taking behavior insights that are generalizable across social platforms and social settings, we focused on encoding information related to user motion and audio, which is logged at 30 Hz. Motion data consisted of the position (i.e., x, y, z) and orientation (i.e., roll, pitch, yaw) of the user’s headset, two controllers, and the ``root’’. The ``root’’ tracks the user’s global position and orientation within the virtual environment, while the headset and controllers are tracked within the coordinate system dictated by the ``root’’. From the tracked motion, we can derive information regarding a user’s egocentric behavior as well as those related to other users. The audio information is recorded both through audio files and a floating-point value between 0 and 1 representing volume.

\subsection{Categories of Turn-Transition Behaviors}
\subsubsection{Pre-processing Speech Events}
We began by identifying all speech events by labeling Inter-Pausal-Units (IPU) \cite{Koiso1998}, defined as the stretches of speech activity by a single speaker. Using the audio tracking data, we determined users as actively speaking when their speech volume is greater than 0.1. To reduce noise from short pauses between speaking activities, we joined adjacent speech events of the same user when the gap between them is within 0.5s. The threshold of 0.5s follows prior literature that found the average pause duration during read speech, interviews and public presentations to be between 0.38 to 0.53 seconds \cite{Liu2022, Campione2002}.

For each session recording, we proceeded to determine the main speaker by first labeling users who are the sole speaker during speech events as the main speaker. When there are multiple speakers, we implemented the following labeling scheme. To start, we eliminated speech events that are completely overlapped by another speech event. Then, we assign the main speaker as the speaker who ends their speech event last. The start time of the new main speaker is marked as the moment the previous main speaker finished speaking. We used Python for extracting IPUs and assigning main speakers. 

\subsubsection{Defining and Labeling Turn-transition Behaviors}
\begin{figure*}[t]
 \centering 
 \includegraphics[width=0.98\textwidth,keepaspectratio]{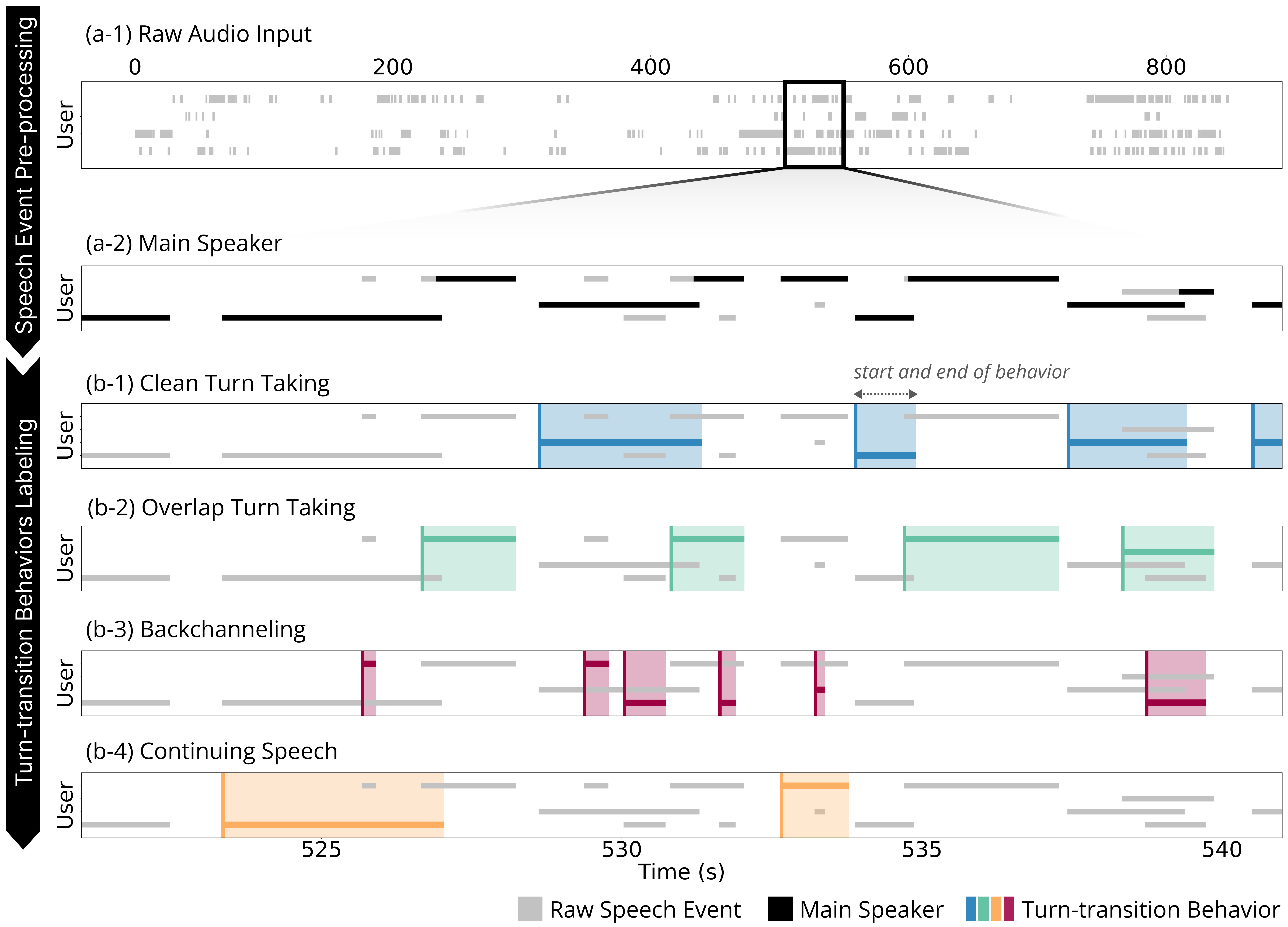}
 \caption{Categorization of Turn-Transition Behaviors based on Example Raw Audio Input. Each row represents a unique user in the VR group discussion session. (a) shows example of speech event pre-processing and (b) shows subsequent behavioral labeling. In (a-2), the main speaker is labeled with black-colored segments. In (b-1)--(b-4), the start of each turn-transition behavior is denoted by a vertical line and its duration shaded in with the color associated with the turn-transition behavior. The speech utterances plotted are based on a recorded VR session and are further edited to improve visual clarity (e.g., adjusting speech events in close proximity).}
 \label{fig:speechCategories}
\end{figure*}

From the labeled main speakers, we extracted four types of turn-transition behaviors following closely the categorizations used by Jokinen et al. \cite{Jokinen2013}: clean turn taking, overlap turn taking, backchanneling, and continuing speech. These categories provide a formal framework for us to examine turn-taking behaviors. Upon obtaining the labels, we removed data points associated with speech events shorter than 323 milliseconds (i.e., the average duration for enunciating fast words found in \cite{Sommers2006}) to filter out noise. We used Python for processing the audio input. Figure \ref{fig:speechCategories} shows categorization of each of the categories based on an example audio input. Table \ref{tab: label summary stats} shows the categories' summary statistics.\footnote{A repository with code used for labeling the four categories is available at \url{github.com/pwang1230/turntaking-wang-2025}.}

\begin{table*}
  \small
  \begin{threeparttable}[b]
  \caption{Summary Statistics of Turn-Transition Categories. The table presents the count of labels for each category by week and their sum across all four weeks. In parenthesis, we report the frequency of categories, where the unit is occurrences per minute across all sessions. Notably, the frequencies and counts of clean turn taking and continuing speech exceed that of overlap turn taking and backchanneling.}
  \label{tab: label summary stats}
  \begin{tabular}{p{0.2\linewidth}ccccc}

    & Week 1 & Week 2 & Week 3 & Week 4 & \textbf{Total}\\
    \midrule
     
    Clean Turn Taking & 1675 (4.42)& 2386 (4.71) & 1712 (5.12) & 2149 (4.87) & \textbf{7922 (4.77)}\\
    Overlap Turn Taking & 622 (1.64)& 939 (1.86) & 771 (2.30) & 904 (2.05)& \textbf{3236 (1.95)}\\
    Backchanneling & 571 (1.51)& 947 (1.87)& 754 (2.25) & 771 (1.75) & \textbf{3043 (1.83)}\\
    Continuing Speech & 1866 (1.12) & 2446 (4.83) & 1600 (4.78) & 2292 (5.20) & \textbf{8204 (4.94)}\\
     
  \end{tabular}
  \end{threeparttable}
\end{table*}

\begin{itemize}

  \item \textbf{Clean Turn Taking}. Clean turn taking occurs whenever the main speaker has changed and that there was no overlapping between the previous and new speaker turns. The start of a clean turn taking event was marked as the beginning of the new speaker’s speech event. Figure \ref{fig:speechCategories}b-1 shows examples of these behaviors.

  \item \textbf{Overlap Turn Taking}. As shown in Figure \ref{fig:speechCategories}b-2, overlap turn taking occurs whenever the main speaker changes and the two speech events overlap. In other words, the start of the speech event of the new speaker precedes the end of the previous speech event. We again marked the start of a turn taking event as the beginning of the new speaker’s speech event.

  \item \textbf{Backchanneling}. We determined backchanneling by checking for instances when a non-main speaker began and ended their speech event without ever being the main speaker. The start of a backchanneling event is labeled as the start of the speech event of the non-main speaker. We show examples of backchanneling in Figure \ref{fig:speechCategories}b-3. One noteworthy scenario to this labeling scheme is when the non-main speaker's speech event covers a change in the main speaker. As shown in Figure \ref{fig:speechCategories}b-3, specifically the final backchanneling event by the bottom user, the main speaker changed during the speech event. Since the bottom user was never labeled the main speaker during their speech event, we labeled this instance as backchanneling.

  \item \textbf{Continuing Speech}. Continuing speech is similar to the scattered contribution category proposed by Jokinen et al. \cite{Jokinen2013}. In their formulation, scattered contributions are defined as two adjacent speech instances of the same speaker being spaced less than 200ms apart, while those that are further apart are considered sequences of turns by the same speaker. Since we are interested in understanding how and when turn transitions take place (i.e., changes in speakers), we combined these two scenarios under the broader category of continuing speech. More concretely, we label continuing speech by first locating instances of adjacent speech events of the same user and then marking the start of the later speech event as the beginning of the turn-transition behavior. Figure \ref{fig:speechCategories}b-4 shows instances of continuing speech behaviors.

\end{itemize}

\subsection{Feature Selection}
We formulated features based on prior literature. As reviewed in Section \ref{Sec: Related work}, user motion, individual and group differences, and verbal behaviors are predictive of user intentions and social dynamics. Past works have also shown that information relating to the current speaker \cite{Chen2024} and their relationship with listeners \cite{Ishii2016Gaze, Ishii2017Head} offer insights to speech behaviors. Importantly, although our dataset varied the virtual spatial context, we chose to not encode these differences because they were manipulated at two distinct levels (i.e., high and low for ceiling height, large and small for amount of visible space). Encoding this study-driven parametrization into predictive models will not generalize to new spatial contexts. Instead, we treat these spatial variations as stimulus sampling to enhance external validity \cite{Wells1999, Jun2022}.

Formally, we extracted features related to listeners, previous speakers, the dyadic relationship between listeners and speakers, their relationship with the group, and verbal behaviors. We further encoded features related to two users: a user of interest which we refer to as the \emph{main user}, and the previous speaker, which we refer to as the \emph{reference user}. When extracting motion-related features, we examined the tracking data of 1-second windows at moments prior to the start of turn-transition behaviors. The 1-second window size is determined based on piloting and aligns with past literature predicting speaking intentions \cite{Chen2024}. Table \ref{tab: feature selection} presents a summary of extracted features.

\begin{table*}
  \footnotesize
  
  \begin{threeparttable}[b]
  \caption{Summary of Features Selected for Prediction Tasks. We present features by their groups, namely those related to speech, individual and group differences, egocentric motion, dyadic relationship, and group relationship.}
  \label{tab: feature selection}
  \renewcommand{\arraystretch}{1.3}
  \begin{tabular}{p{0.21\linewidth}p{0.43\linewidth}p{0.25\linewidth}}
  
    Feature Group & Description of Features & Number of Features\tnote{1}\\
    \midrule
        
    Speech-Related \cite{Parker1988}& Prior speech sequences & 10 preceding speaker indices\\
    & Number of speaking turns before the main user’s last speech event& 1\\
    & Time before the end of the main user's last speech event& 1\\
    & Whether the main user has spoken before & 1\\
    
    \multirow{2}{=}{Individual- and Group- Level Differences \cite{Ramsay1968, Ivanov2011, Aran2013, Mast2002, Hadley2021, Lepri2010}} &  Main user big-5 personality & 5\\
    & Reference user big-5 personality & 5\\
    & Big-5 personality averaged across all users & 5\\
    & Group size & 1\\

    Egocentric Motion\tnote{2} \newline \cite{Miller2020, Miller2023VR, Chen2024, Ishii2017Head, Aran2013} & Position and orientation of a user's headset and two hand-held controller, where the user is either the main user or reference user & 3 tracked points $\times$ 6 DOFs $\times$ 3 summary stats. $\times$ 2 meas. $\times$ 2 users = 216 \\

    \multirow{2}{=}{Dyadic Relationship \cite{Chen2024, Jokinen2013, Ishii2016Gaze, Lee2023}} & Direct visual attention from the main user to the reference user & 3 summary stats. $\times$ 2 meas. = 6\\
    & Direct visual attention from the reference user to the main user & 3 summary stats. $\times$ 2 meas. = 6\\
    & Interpersonal distance between the main user and reference user & 3 summary stats. $\times$ 2 meas. = 6\\
    & Visual shared space calculated between the main user and reference user & 3 summary stats. $\times$ 3 dists. $\times$ 2 meas. = 18\\
    
    Group Relationship \newline \cite{Chen2024, Jokinen2013, Ishii2016Gaze, Lee2023}& Direct visual attention from a user to all remaining users, where the user is either the main user or reference user & 3 summary stats. $\times$ 2 meas. $\times$ 2 users = 12\\
    & Direct visual attention from all remaining users to a user, where the user is either the main user or reference user & 3 summary stats. $\times$ 2 meas. $\times$ 2 users = 12\\
    & Interpersonal distance between a user and all remaining users, where the user is either the main user or reference user & 3 summary stats. $\times$ 2 meas. $\times$ 2 users = 12\\
    & Visual shared space calculated between a user and all remaining users, where the user is either the main user or reference user & 3 summary stats. $\times$ 3 dists. $\times$ 2 meas. $\times$ 2 users = 36\\
     
  \midrule
\end{tabular}

     \begin{tablenotes}
     \item[1] 3 Tracked points refer to the headset and two controllers; 6 degrees of freedom refer to x, y, and z position and the roll, pitch, and yaw orientation; 3 summary statistics refer to the minimum, maximum, and average across the examined time period; 2 measurements refer to the raw value and first order derivative (i.e., velocity); 3 distances refer to 1, 5, and 10 meters. DOFs = degrees of freedom; stats. = statistics; meas. = measurements; dists. = distances.
     \item[2] Due to the body-space transformation and centering operations, the total number of non-trivial features for egocentric motion is 198.
     \end{tablenotes}
    
  \end{threeparttable}
\end{table*}

\subsubsection{Speech-Related Features}

The first feature group is related to speech behaviors, as they were shown to be predictive of who the next speaker is \cite{Parker1988}. Following Parker \cite{Parker1988}, we encoded speech sequences leading up to the start of the examined window. For this, we begin at the timestamp at the start of the examined window, and trace backwards each time the main speaker changes, at which point we increment the turn index by 1. If the main speaker for a given turn is the main user, we assign the feature value at the corresponding turn index ``u’’. If we encounter a new speaker at a preceding turn, we assign the feature value a new symbol representing that user (e.g., ``a’’, ``b’’, ``c’’). For turns with a speaker who we had already created a symbol for, we assigned their symbol to the feature value. In our setup, we encoded 10 preceding speaker indices, assigning ``NA’’ when there were no more changes in main speakers preceding the last encoded turn. In practice, we processed the speech sequence features at each turn index using one-hot encoding.

We introduced three additional features to describe the main user’s verbal behaviors: (1) the number of speaking turns before the main speaker’s last speech event, (2) the duration of time before the end of the main user’s last speech event, and (3) whether the main user has spoken before during the current session. Deriving these features for the reference user was unnecessary since that user is typically set to the previous speaker.

\subsubsection{Individual- and Group- Level Features}

Drawing from works showing correlations between speech behaviors and personalities and group compositions \cite{Ramsay1968, Ivanov2011, Aran2013, Mast2002, Hadley2021}, we formulated features based on individual and group characteristics. Specifically, we encoded the Big-5 personality traits \cite{Gosling2003Big5} of the main and reference users, and group average. The authors of the dataset measured Big-5 personality using the ten item personality measure (TIPI) \cite{Gosling2003Big5}, with each personality trait ranging between 1--7 and calculated as the mean of two items.

\subsubsection{Egocentric Motion Features} \label{sec: egocentric features}

Egocentric motion features capture the pose and motion of individuals. For this, we extracted egocentric motion features from both the main and reference users. For each user, we extracted the position (i.e., x, y, z) and orientation (i.e., pitch, yaw, roll) of a user’s headset and two controllers by both their raw values and velocities. From the raw values and velocities across the 1-second window, we summarized each of the 6 degrees of freedom by their average, minimum, and maximum. We transformed the features corresponding to the raw values into the body-space coordinate system proposed in Miller et al. \cite{Miller2023Identification} and centered the horizontal coordinates (i.e., x, z) such that the individual’s head position is at the origin of the horizontal plane, as opposed to their global coordinates in the virtual space. The rationale behind this transformation is to encode the egocentric poses by mapping the raw coordinates and orientations to a coordinate system based on the individual’s forward head direction. This allows the features to encode poses even when individuals move around physical or virtual environments. 

When calculating the velocity of positions and orientations, we derived velocities from raw values after transforming the original values into the body-space coordinates without the centering operation. Not centering the coordinates based on the head position allows the features to retain information of the user’s head position velocities in the horizontal plane. When encoding head yaw, we did not calculate its velocity based on the transformed values but instead on the yaw angles prior to any transformations. This procedure preserves information on how a user rotates their head in the yaw axis, which body-space transformation loses by reorienting based on head yaw.


\subsubsection{Dyadic Relationship Features}

The next feature group describes the dynamics between the main and reference users. Similar to prior work \cite{Ishii2016Gaze, Chen2024, Jokinen2013}, we extracted the direct visual attention (i.e., gaze direction approximated using head orientation) from the main user to the reference user, and vice versa. Specifically, we determined user A’s direct visual attention towards user B by calculating the angle between user A’s forward head orientation in the yaw axis and the vector pointing from user A’s head position to user B’s head position in the horizontal plane. Similar to the egocentric motion features, we derived the raw values and velocities across the 1-second window and summarized them by their average, minimum, and maximum. Since users moved around virtual environments in the dataset, we also encoded interpersonal distance as the distance between two users’ head position in the horizontal plane. 

One characteristic that direct visual attention fails to capture is whether two users are facing the same direction. For example, if both users' head orientations are at 90 degrees from the other user, they could be facing the same direction or have their backs against one another. These two scenarios can entail different dynamics as users facing the same direction are likely looking at similar parts of the environment. We therefore introduced a measurement of visual shared space quantifying how much users' visual field of views overlap. Concretely, we first drew isosceles triangles with the vertex connecting the two equal sides of length $vs_l$ located at the user’s head and oriented in the head’s forward direction in the horizontal plane. The angle between the two sides is set to the horizontal field of view of the Meta Quest 2 headset (104 degrees). The raw value of the visual shared space between two users at distance $l$ is defined as the amount of spatial overlap in $m^2$ of the two triangles drawn with $vs_l=l$. In simpler terms, this measurement estimates how much virtual space the two users share in their field of views, assuming that their visual attention stretches $vs_l$ meters. For our setup, we extracted values for $vs_l$ at 1, 5, and 10 meters.

Using the raw values calculated for direct visual attention, interpersonal distance and visual shared space, we extracted the raw values and their changes (i.e., velocity) over the sampled windows and summarized the two measurements into their average, minimum, and maximum.

\subsubsection{Group Relationship Features}

Finally, we encoded features between a user of interest (i.e., main or reference user) with the group. Specifically, we extracted (1) the dyadic direct visual attention angle from each user of interest to each group member and (2) those from each group member to that user. Similar to the dyadic relationship features, we calculated the interpersonal distance between each of the two users with the rest of the group. Finally, we derived the visual shared space between each dyadic pair between a user of interest and the rest of the group at $vs_l$ = 1, 5, 10 meters. From the raw values extracted for direct visual attention, interpersonal distance, and visual shared space across the sampled window, we derived the average value and velocity for each dyadic pair, grouped them by whether they are related to the main or reference user, and finally summarized them by their average, minimum, and maximum.

\subsection{Turn-Taking Behavior Prediction Tasks} \label{sec: prediction tasks}

Drawing from past literature on predicting speech behaviors \cite{Ishii2016Gaze, Ishii2017Head, Chen2024, Lee2023, Parker1988}, we focused on three prediction tasks, which capture the ``what’’, ``who’’, and ``when’’ of turn-taking behaviors. 

\begin{itemize}

  \item \textbf{Turn Taking vs. Continuing Speech}. The ``what’’ poses the question of whether we can predict the type of turn-transition behavior. In particular, in line with past research \cite{Ishii2016Gaze, Lee2023, Jokinen2013}, can we distinguish whether a turn-transition behavior will be a turn-transition to a new speaker (i.e., turn taking), or a continuing speech from the previous speaker (i.e., turn keeping)?

  \item \textbf{Next Speaker Prediction}. The ``who’’ focuses on predicting who the new speaker is prior to the start of a turn-taking behavior (i.e., clean turn taking, overlap turn taking). Next speaker prediction is another common task used in modeling speech behavior \cite{Ishii2016Gaze, Parker1988, Ishii2016Respiration, Ishii2016Mouth}.

  \item \textbf{Timing of Turn-Taking Behaviors}. To investigate the ``when’’ of turn-taking behaviors, we investigate whether we can predict when the next speaker will speak, a task that frequented past literature \cite{Ishii2016Gaze, Ishii2016Respiration, Ishii2017Head}. Specifically, can we differentiate between moments associated with a new speaker right before a turn taking event and those sampled before these moments?

\end{itemize}

\subsection{Machine Learning Models} \label{sec: machine learning models}

To study \textbf{RQ1}, we predicted the three tasks outlined in Section \ref{sec: prediction tasks} by first extracting the features in the 1-second windows prior to the start of the turn-transition behaviors. For each task, we built and compared performance across four predictive models commonly used to predict individual characteristics and group dynamics \cite{Cao2021, Miller2023Identification, Moore2021}: logistic regression, multi-level perceptron (MLP) classifier, random forest classifier, and gradient boosting classifier. We formulated each task as a binary classification task and measured prediction performance using the area under the curve (AUC) of the receiver operating characteristics curve. We implemented all models using Python’s scikit-learn library \cite{ScikitLearn}. 

For evaluation, we employed cross-validation similar to \cite{Frommel2020, Cao2021}\footnote{We conducted an additional analysis that focused on evaluating model performance across sessions, groups, and weeks. Since the results largely align with our cross-validation evaluations, we include this analysis in Appendix \ref{apx: Additional Results on Model Predictions}.}. To start, we partitioned the data into training and testing and standardized all continuous features. For measuring accuracy, we reported the performance of models trained on 90\% of the data and evaluated on the remaining 10\% (i.e., k-fold $\approx$ 10). Specifically, we calculated the averages and standard errors of the AUC evaluated on the testing data across all folds. As a benchmark measurement, we partitioned the data such that we trained on data from 90\% of the sessions and tested was on the remaining 10\% (Session\textsubscript{cv}). 

We report two additional sets of results evaluating the robustness of models on unseen groups, activities, and weeks (\textbf{RQ2}). First, we measured performance on unseen groups by partitioning the data such that we trained on data from 90\% of groups and tested on the remaining 10\% (Group\textsubscript{cv}). We then quantified performance across unseen weeks using models trained on all but one week’s data and tested on the remaining week (Week\textsubscript{cv}). Finally, we report performance on the final week after training models on data from the first three weeks. As group activities differed weekly, our evaluation metric for unseen activities is the same as that for unseen weeks.

\section{Results}

In this section, we report on model performance  (\textbf{RQ1}) and their robustness (\textbf{RQ2}) for the prediction tasks outlined in Section \ref{sec: prediction tasks}. Then, we present analyses investigating feature importance and how the features are related to model predictions (\textbf{RQ3}).

\subsection{Predicting Turn-Taking Behaviors} \label{sec: task prediction results}
\subsubsection{Differentiating between Turn Taking and Continuing Speech} \label{res: task1}

  
     
     

\begin{table*}
  \small
  \begin{threeparttable}[b]
  \caption{Model Performance (measured as the AUC of the ROC curve) on Predicting Turn-Taking Behavior vs. Continuing Speech. For metrics using cross validation, which we denote using the subscript cv, we report the average and standard error across all folds. An AUC of 0.50 means that the model's ability to distinguish between positive and negative samples is no better than random chance. Bolded numbers denote best performance by metric.}
  \label{tab: task 1 performance}
  \begin{tabular}{p{0.26\linewidth}cccc}
    & \multicolumn{4}{c}{Performance Metrics}\\ 
    \cmidrule{2-5}
  
    Prediction Model& Session\textsubscript{cv}& Group\textsubscript{cv} & Week\textsubscript{cv} & Week 4\\
    \midrule
     
    Logistic Regression & 0.71 (0.01) & 0.69 (0.01) & 0.70 (0.00) & 0.69\\
    MLP Classifier & 0.72 (0.01) & 0.71 (0.01) & 0.71 (0.00) & 0.71\\
    Random Forest Classifier & 0.77 (0.00) & 0.75 (0.01) & 0.76 (0.00) & 0.76\\
    Gradient Boosting Classifier & \textbf{0.78 (0.00)} & \textbf{0.77 (0.01)} & \textbf{0.78 (0.00)} & \textbf{0.78}\\
     
  \end{tabular}
  \end{threeparttable}
\end{table*}

We built models for predicting whether a speech event was going to be a turn transition or one where the previous speaker will continue to speak. We compared clean turn taking, and not overlap turn taking, to continuing speech as there are by definition pauses before both types of turn-transition behaviors. We defined positive samples as those corresponding to the beginning of turn-taking behaviors and negative samples as those corresponding to the start of continuing speech behaviors. We chose the upcoming speaker as a main user for positive samples and a randomly selected user who is not the previous speaker for negative samples. For both positive and negative samples, the reference user was the previous speaker. We randomly sampled the dataset to maintain an equal number of positive and negative samples, which yielded 15844 samples.

We present model performance in Table \ref{tab: task 1 performance}. Notably, gradient boosting classifiers outperformed other models across all performance metrics, achieving a benchmark accuracy (i.e., Session\textsubscript{cv}) of 0.78 AUC, and 0.77--0.78 AUC across the three remaining metrics. Random forest classifiers achieved the second highest accuracy on the benchmark metric with a 0.77 AUC, followed by the MLP classifier at 0.72 AUC, and the logistic regression at 0.71 AUC. When evaluating on unseen groups, performance accuracies were slightly lower but generally robust across all four models. Compared to the benchmark metric, models achieve similar accuracies on unseen weeks and activities.

\subsubsection{Predicting the Next Speaker} \label{res: task2}

\begin{table*}
  \small
  \begin{threeparttable}[b]
  \caption{Model Performance (measured as the AUC of the ROC curve) on Next Speaker Prediction. For metrics using cross validation, which we denote using the subscript cv, we report the average and standard error across all folds. An AUC of 0.50 means that the model's ability to distinguish between positive and negative samples is no better than random chance. Bolded numbers denote best performance by metric.}
  \label{tab: task 2 performance}
  \begin{tabular}{p{0.26\linewidth}cccc}
    & \multicolumn{4}{c}{Performance Metrics}\\ 
    \cmidrule{2-5}
  
    Prediction Model& Session\textsubscript{cv}& Group\textsubscript{cv} & Week\textsubscript{cv} & Week 4\\
    \midrule
     
    Logistic Regression & 0.73 (0.01) & 0.72 (0.01) & 0.73 (0.01) & 0.75\\
    MLP Classifier & 0.72 (0.01) & 0.71 (0.01) & 0.72 (0.01) & 0.74\\
    Random Forest Classifier & 0.75 (0.01) & 0.74 (0.01) & 0.74 (0.00) & 0.75\\
    Gradient Boosting Classifier & \textbf{0.77 (0.01)} & \textbf{0.75 (0.01)} & \textbf{0.77 (0.01)} & \textbf{0.78}\\
     
  \end{tabular}
  \end{threeparttable}
\end{table*}

We predicted the upcoming speaker of a turn-taking event. Specifically, we aggregated positive samples by extracting those corresponding to moments at the beginning of turn-taking events (i.e., clean turn taking, overlap turn taking), where the main user is set to the upcoming speaker. We then collected negative samples at the same moments, but with the main speaker set to a user who is neither the upcoming nor the previous speaker. The reference user is set to the previous speaker for all samples. We randomly resampled negative labels to balance positive and negative samples, yielding 22316 samples.

Table \ref{tab: task 2 performance} summarizes our results. Gradient boosting classifier achieved the highest accuracies across all four performance metrics, with an AUC of 0.77 on the benchmark metric, followed by the random forest classifier at 0.75 AUC, the logistic regression at 0.73 AUC, and finally the MLP classifier at 0.72 AUC. All four prediction models were robust when tested on unseen groups and weeks, with the performance on unseen groups again being lower but comparable to the benchmark evaluations.

\subsubsection{Predicting the Timing of Turn Taking} \label{res: task3}

\begin{table*}
  \small
  \begin{threeparttable}[b]
  \caption{Model Performance (measured as the AUC of the ROC curve) on the Timing of Turn Taking. For metrics using cross validation, which we denote using the subscript cv, we report the average and standard error across all folds. An AUC of 0.50 means that the model's ability to distinguish between positive and negative samples is no better than random chance. Bolded numbers denote best performance by metric.}
  \label{tab: task 3 performance}
  \begin{tabular}{p{0.26\linewidth}cccc}
    & \multicolumn{4}{c}{Performance Metrics}\\ 
    \cmidrule{2-5}
  
    Prediction Model& Session\textsubscript{cv}& Group\textsubscript{cv} & Week\textsubscript{cv} & Week 4\\
    \midrule
     
    Logistic Regression & 0.61 (0.00) & 0.61 (0.00) & 0.61 (0.00) & 0.60\\
    MLP Classifier & 0.63 (0.00) & 0.63 (0.01) & 0.63 (0.00) & 0.61\\
    Random Forest Classifier & 0.68 (0.00) & 0.67 (0.01) & 0.67 (0.01) & 0.66\\
    Gradient Boosting Classifier & \textbf{0.72 (0.00)} & \textbf{0.71 (0.01)} & \textbf{0.71 (0.00)} & \textbf{0.71}\\
     
  \end{tabular}
  \end{threeparttable}
\end{table*}

Finally, we aimed to understand whether we can differentiate between moments associated with an upcoming speaker right before a turn taking event and those sampled prior to these moments. We first collected positive samples by extracting samples corresponding to moments before the start of turn-taking events (i.e., clean turn taking, overlap turn taking), with the main user set to the upcoming speaker. For negative samples, we aggregated samples corresponding to moments that were prior to the 1-second window immediately before the start of turn-taking event. For this, we began by sampling moments at windows starting 2, 4, 6, 8, 10, and 12 seconds prior to the start of turn transition events. We then filtered out data samples for which any speech event had occurred between the earlier moment and the start of the turn transition event for the upcoming speaker. In other words, the negative samples consisted only of moments when the upcoming speaker is not preparing to speak. This sampling technique is similar to Chen et al. \cite{Chen2024} as we sampled through a sliding window. For both positive and negative samples, the reference user is set to the previous speaker at the sampled moment. We randomly resampled the aggregated data to ensure an equal number of positive and negative samples, resulting in 22316 samples.

We detail performance results in Table \ref{tab: task 3 performance}, which were lower than those from previous tasks. The best performing model, the gradient boosting classifier, achieved an accuracy of 0.72 AUC on the benchmark metric, while the logistic regression model recorded the lowest accuracy of 0.61 AUC on the same metric. The random forest and MLP classifiers had benchmark accuracies of 0.68 and 0.63 AUC, respectively. Performance on unseen groups, activities, and weeks were robust, with models reporting slightly lower but similarly accuracies on other cross validation evaluations.

\subsection{Feature Importance Analysis}
\subsubsection{Mean Decrease in Accuracy Based on Gradient Boosting Classifiers}
\begin{figure*}[t]
 \centering 
 \includegraphics[width=0.88\textwidth,keepaspectratio]{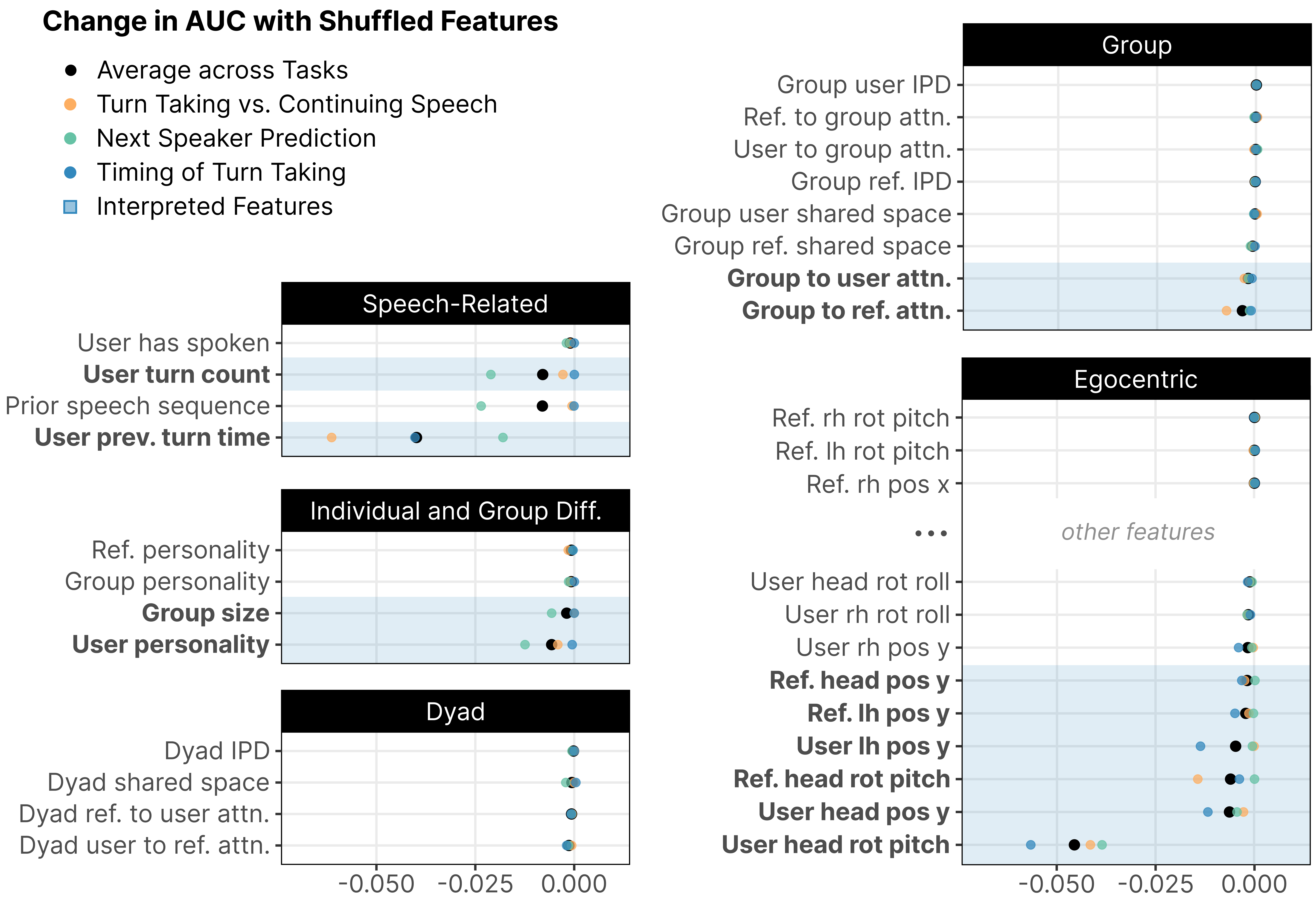}
 \caption{Feature Importance Analysis. Each panel reports the change in AUC of the ROC after shuffling features across the three prediction tasks for a given feature group. Black dots denote the average drop in accuracy across three tasks. Features within each panel are ordered in increasing average drop in accuracy from top to bottom. Lower values (i.e., greater decrease in accuracy) represents higher MDA. For clearer presentation, we did not plot egocentric features with moderate importance. Bolded features and their corresponding rows with light blue backgrounds indicate features that are further interpreted in Section \ref{sec: feature interpretation}. Diff. = differences; Ref. = reference user; prev. = previous; IPD = interpersonal distance; attn. = visual attention; lh = left hand; rh = right hand.}
 \label{fig: feature importance}
\end{figure*}

To understand how features are related to model performance and predictions \textbf{(RQ3)}, we first computed the mean decrease in accuracy (MDA) for the top performing model (i.e., gradient boosting classifier), a technique used for probing feature importance in machine learning models \cite{Miller2020, Han2016MDA, Nicodemus2011MDA}. Specifically, we organized features based on their feature groups and definition, for example splitting up egocentric motion features by the user, tracked points, and axis of motion. For each task, we built models used to compute the benchmark cross validation performance (i.e., Session\textsubscript{cv}) and evaluated them on each testing set with a particular set of features randomly shuffled. This procedure breaks up any relationship the set of features may have with the model prediction. We then calculated the average change in the testing AUC across all folds and present them in Figure \ref{fig: feature importance}. We omitted plotting egocentric features with moderate importance, but included all features in Appendix \ref{apx: full feature importance}.

Analysis of speech-related features showed the highest importance on the user’s previous turn time, followed by prior speech sequence, user turn count, and whether the user has spoken earlier in the session. For the main user's previous turn time feature, feature importance for the turn taking vs. continuing speech task was the highest, followed by the timing prediction and next speaker prediction tasks. For features on the main user's turn count and prior speech sequence, they exhibited the highest importance for models built to predict the next speaker, followed by the turn taking vs. continuing speech task, andlastly the timing prediction task. Our results on the individual and group differences features revealed a comparably high importance of the main user personality, followed by group size, group personality, and lastly the reference user personality. 

For egocentric motion features, the top six features were related to the main and reference users' head yaw rotation, head y-axis position, and left hand y-axis position. Within the top six features, feature importance for those related to the main user were in most cases higher for the timing prediction task compared to other tasks. For dyad relationship features, those related to direct visual attention had greater importance than those related to the visual shared space and interpersonal distance, though their change in AUC near zero implies low importance. In terms of group relationship features, features related to the group’s direct visual attention towards the reference and main users exhibited greater importance than the remaining features. 

\subsubsection{Feature Significance Based on Logistic Regression Models}\label{sec: feature significance}
We investigated predictor significance of feature groups using logistic regressions. Benefits of studying linear models include their high interpretability and simplified independence assumption affording well-defined hypothesis tests for evaluating feature significance \cite{Couronne2018}. In contrast, recent methods for interpreting more complex models such as the random forest and gradient boosting classifiers have not addressed evaluations of feature importance through statistical tests \cite{Greenwell2018, Lundberg2017, Lundberg2018}. For this, we grouped features based on their constructs. Then, for each trained logistic regression, we conducted Wald tests to evaluate joint linear hypotheses, specifically testing whether all features related to a given construct significantly contribute to the model. We present the full results in Appendix \ref{apx: stats logistic regression} but highlight general trends for the remainder of this section. We evaluated significance at $\alpha$=.05.

Specifically, all but one group of speech-related features significantly contributed to model prediction across all three tasks ($p$s<.046). For individual and group differences features (i.e., personality, group size), most significantly impacted the tasks of predicting turn taking vs. continuing speech and next speaker prediction ($p$s<.020). In contrast, for individual and group differences features, only features related to the previous speaker's personality ($p$<.001) significantly predicted turn-taking timing. For dyad-related features, visual attention features between the reference and main users significantly predicted the next speaker ($p$s<.001) while the interpersonal distance between the reference and main users was significant in distinguishing between turn taking and continuing speech behaviors ($p$=.004). For features related to groups dynamics, all constructs significantly predicted turn taking vs. continuing speech ($p$s<.017). Group features related to visual attention were also significant in predicting at least one of the two remaining tasks. Finally, for egocentric features, 18 groups of constructs were significant in predicting turn taking vs. continuing speech ($p$s<.035), while only 6 were significant in predicting the next speaker ($p$s<.027). Notably, all 6 were related to the main user's egocentric behavior. For predicting the timing of turn taking, 12 groups of constructs were significant ($p$s<.043).

\subsection{Feature Interpretation Analysis} \label{sec: feature interpretation}

To study \textbf{RQ3}, we analyzed how features with high importance are related to predictions of the best performing models (i.e., gradient boosting classifiers). We selected features within each feature group with high feature importance based on MDA. Notably, we included features deemed insignificant in the Wald-tests in Section \ref{sec: feature significance} for two reasons. First, we interpreted the gradient boosting classifiers and not the logistic regressions. Using MDA from the gradient boosting classifers allowed us to capture insights from features that may lack significance under the assumptions of a linear model but still contribute non-linearly. Additionally, logistic regressions are sensitive to collinearity and can complicate assessments of feature significance \cite{Dormann2013}. Another issue is an overemphasis on features with small effects given a large sample \cite{Sullivan2012}. By leveraging MDA instead, we selected features that best reflected their contribution to the best performing models.

In aggregate, we analyzed 56 constructs. For speech-related features, we examined 2 constructs: the user turn count and user previous turn time. There were 6 constructs related to individual and group differences: main user personalities and group size. 12 were related to the group relationship, specifically those capturing the group’s direct visual attention towards the main user and reference user. The remaining 36 were egocentric motion constructs, namely those describing the head position in the y-axis, rotation in the yaw axis, and left hand position in the y-axis for the main and reference users. We did not interpret individual prior speech sequence features as research has suggested the need to consider the interactions between prior speakers \cite{Parker1988}.

We interpreted features using partial dependence, a common technique used to interpret machine learning models such as random forests \cite{Furlanello2003, Auret2012} and gradient boosting classifiers \cite{Natekin2013, Blagus2017}. By definition, partial dependence measures the marginal effect on model prediction through varying the values to a feature and calculating the average probability estimates after this procedure. In our setup, we chose the gradient boosting classifiers trained on the entire dataset and varied each feature between its [0.05, 0.95] percentile.

\begin{figure*}[t]
 \centering 
 \includegraphics[width=0.9\textwidth,keepaspectratio]{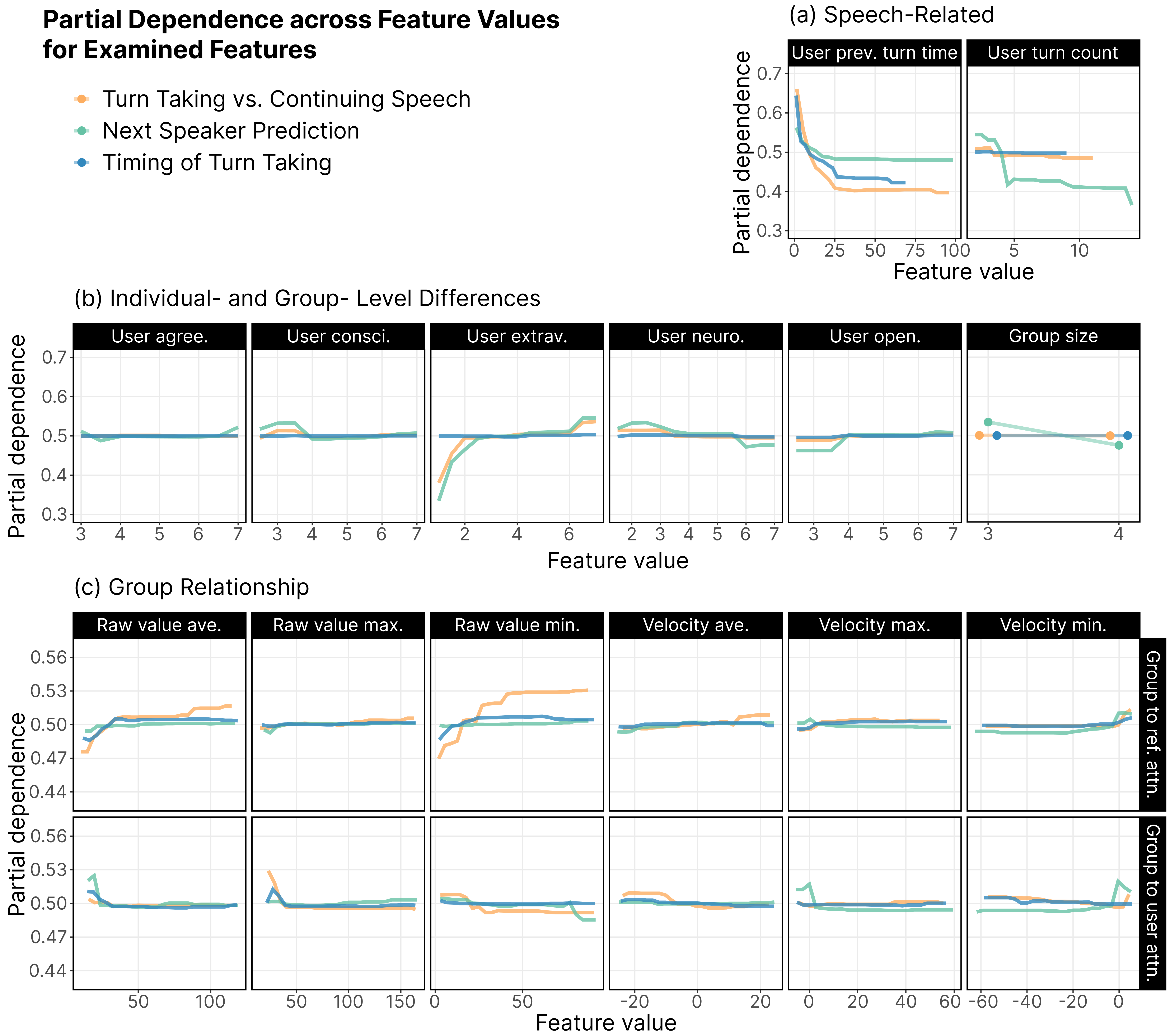}
 \caption{Partial Dependence for Selected Features Related to Speech, Individual- and Group- Level Difference, and Group Relationship. Higher y values denote greater probability estimates for the main user’s speaking intentions. Raw values are quantified in degrees, and velocities in degrees per second. (a--c) show the partial dependence of the selected features, with colors denoting the prediction task. Features related to speech, the main user's extraversion, and raw values of direct visual attention from the group to the main and reference users revealed noticeable changes in the average probability estimates across feature values. Ref. = reference user; ave. = average; max. = maximum; min. = minimum; agree. = agreeableness; consci. = conscientiousness; extrav. = extraversion; neuro. = neuroticism; open. = openness; attn. = visual attention.}
 \label{fig: partial dependencies}
\end{figure*}

\begin{figure*}
 \centering 
 \includegraphics[width=0.9\textwidth,keepaspectratio]{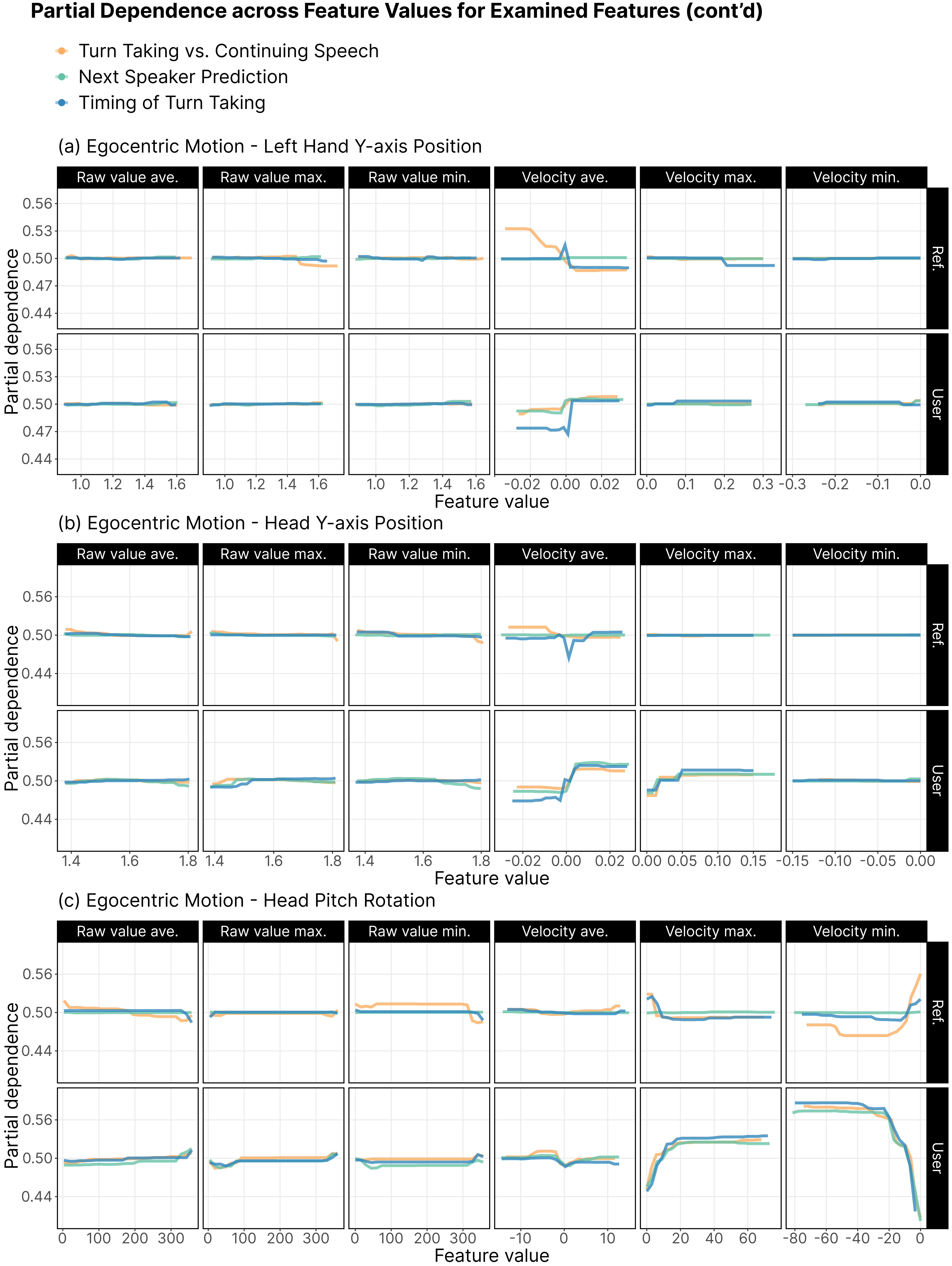}
 \caption{Partial Dependence for Selected Features Related to Egocentric Motion. Higher y values denote greater probability estimates for the main user’s speaking intentions. Raw values are quantified in degrees, and velocities in degrees per second. (a--c) show the partial dependence of the selected features, with colors denoting the prediction task. Notably, results showed the largest visual changes in average probability estimates when varying (1) the average velocity of the users' left hand y-axis position, (2) the maximum and minimum velocities of the users' head pitch rotation, and (3) the average velocity of the users' head y-axis position. Ref. = reference user; ave. = average; max. = maximum; min. = minimum.}
 \label{fig: partial dependencies cont'd}
\end{figure*}

\subsubsection{Speech-Related Features}

Seen in Figure \ref{fig: partial dependencies}a, varying the amount of time and number of turns to the main user’s previous turn revealed large effects on the probability estimates. Specifically, the results suggested that listeners who had spoken more recently are more likely to engage in turn-taking behaviors and be the next speaker (i.e., predicting turn taking vs. continuing speech and the next speaker). The models also predicted that listeners are more likely to speak at a time closer to their previous speech activity. Our results on the user turn count feature revealed for the next speaker prediction task that listeners whose last speaking event occurred at earlier turns are less likely to become the next speaker.

\subsubsection{Individual- and Group- Level Differences}

Shown in Figure \ref{fig: partial dependencies}b, for predicting the next speaker and differentiating between turn taking and continuing speech, listeners with higher values of extraversion are more likely to be the next speaker. There was little variation in probabilities estimates when varying main user personalities for predicting the timing of turn-taking. This suggests that listeners with different levels of Big-5 personalities do not exhibit different patterns in \emph{when} they decide to speak. There were weaker but noticeable relationships between the probabilities estimates and the listener’s consciousness, neuroticism, and openness. For example, for the next speaker prediction task, the model predicted that listeners reporting a lower level of openness were less likely to be the next speaker.

\subsubsection{Group Relationship}

The most salient relationship in the partial dependence plots regarding group relationships (Figure \ref{fig: partial dependencies}c) were those related to the group’s direct visual attention towards the previous speaker. Notably, the models predicted that a new speaker is more likely to take over a previous turn when listeners are looking away from the previous speaker (i.e., larger minimum and average value for the direct visual attention angle from the group to the reference user). There is a similar but weaker trend when predicting the timing of turn taking. 

The partial dependence plots for the group’s direct visual attention towards a listener revealed the opposite directionality between prediction estimates and feature values, where a listener is predicted to be more likely to take over a turn when the remaining group is looking more directly at them (i.e., small maximum and average value for the direct visual attention angle from the group to the main user). This pattern is, however, less pronounced than the greater variability in probability estimates exhibited for direct visual attention measurements from the group to the previous speaker. 

\subsubsection{Egocentric Motion}

Figure \ref{fig: partial dependencies cont'd} presents our findings on egocentric motion features. Shown in Figure \ref{fig: partial dependencies cont'd}a, the feature with the greatest variation in partial dependence for the \textbf{left hand y-axis position} was the average velocities for both the listener and previous speaker. Notably, the model used for differentiating between turn taking and continuing speech predicted that previous speakers who move their left hand upward in the vertical axis (i.e., large positive values in the average y-axis velocity) are more likely to continue speaking. For listeners, models for all three tasks predicted that listeners are more likely to speak in moments right after they move their left hand upward (i.e., large positive values in the average y-axis velocity).

For egocentric features related to \textbf{head y-axis position}, the salient trends were related to velocities (Figure \ref{fig: partial dependencies cont'd}b). To start, the model for the turn-taking timing task predicted that a listener is more likely to begin speaking if the previous speaker’s head had on average not moved a substantial amount in either direction in the vertical axis. This was not the case for models for the other two tasks. On the contrary, all three models predicted that a listener is more likely to begin speaking if their head had on average moved upward in the y-axis (i.e., large positive values for the average y-axis velocity). Relatedly, a listener is predicted to more likely speak if their head had at any point in the sampled window moved quickly upward in the vertical axis (i.e., large positive values for the maximum y-axis velocity).

Finally, for features related to \textbf{head pitch rotation}, we found distinctive patterns in maximum and minimum velocities of the pitch rotations for the main and reference users (Figure \ref{fig: partial dependencies cont'd}c). We interpret this as the models finding the extremes of an individual’s head pitch rotation velocities across the 1-second window more informative than their averages. There were no noticeable trends in partial dependence for the previous speaker’s head pitch rotation on the next speaker prediction task. Models for the remaining two tasks predicted that a listener is less likely to speak if the previous speaker has at any point extensively rotated their head upward, which corresponds to small negative values for the minimum pitch velocity. The models also predicted that a listener is less likely to speak if the previous speaker has at any moment within the 1-second window extensively rotated their head downward (i.e., large positive values for the maximum pitch velocity). The trends in partial dependence on the listener’s head pitch rotation were opposite to that of the previous speaker. Namely, all three models predicted a listener is more likely to speak when they have engaged in substantial downward head rotation (i.e., large positive values for the maximum pitch velocity) or upward head rotation (i.e., small negative values for the minimum pitch velocity).

\section{Discussion}

\subsection{Practical Implications of Turn-Taking Prediction}

\subsubsection{Model Selection and Performance}
\label{sec: model selection}
With regards to \textbf{RQ1}, we found that the gradient boosting classifiers achieved consistently the best performance on all tasks across performance metrics. This suggests that the relationship between the features and outcome variable is non-linear and that interactions between features are likely relevant for prediction. Specifically, the models obtained accuracies of 0.75--0.78 AUC on the tasks of identifying the new speaker and distinguishing between turn transitions and continuing speech, and had poorer performance on predicting when a new user will begin speaking with accuracies of 0.71--0.72 AUC. These accuracies are considerably higher than prediction by chance, are considered to have excellent discrimination between positive and negative samples \cite{Yang2017}, and comparable to prior research \cite{Chen2024}. 

As AUC represents the probability that a random positive sample is rated higher than a random negative sample \cite{Bradley1997}, there remains room for prediction improvements, albeit limited by the complexity inherent to social dynamics and VR \cite{Degutyte2021, Maloney2020}. To further explore the practical implications of turn-taking predictions, we present an additional analysis in Appendix \ref{apx: AB model resutls}. In it, we consider the task of predicting continuing speech and turn-taking events and extended the binary classification problem into a multiclass one. This analysis yielded comparable findings and found that three-person groups yielded higher prediction accuracy than four-person ones. One possible explanation for the lower performance in larger groups is the greater unpredictability in larger-group social dynamics. While there were more four-person discussions than three-person ones in our dataset, greater complexity in social dynamics may warrant more data for larger group sizes. When modeling and deploying comparable models in real-world settings, our findings here suggest that practitioners should sample diverse group sizes during data collection, exercise caution when predicting turn-taking behaviors for groups whose sizes differ from the training data, and not assume higher prediction accuracies in smaller groups will translate directly to larger groups.

Additionally, our model accuracies suggest that near-term uses cases should prioritize applications that either 1) are minimally impacted by potential false labels or 2) build off insights from feature importance and interpretation. Given the lower performance in predicting turn taking timing, practitioners should build tools that leverage predictions for who the next speaker is, as opposed to when they will begin speaking. 

Importantly, overall accuracy alone does not capture the full picture to model performance, for generalizability is critical for practical deployments. In our work, we showed that predictive performance is robust over time and across unseen groups and activities (\textbf{RQ2}), with accuracies for evaluations based on unseen groups being slightly lower but comparable to benchmark metrics. Considering also the diverse group activities and virtual mobility allowed in our dataset, these results suggest that practitioners can expect robustness across time and unseen activities and groups when training and deploying predictive models using similar approaches. One possible explanation for the robustness is the high number of groups and wide range of activities seen during training allowing models to extrapolate generalizable insights.

\subsubsection{Application of Findings}\label{sec: applications} Our findings offer insights for practical applications. For one, for educators and organizers facilitating group discussions, having a tool for predicting next speakers can help with moderation (e.g., smoother transition between speakers). As these facilitators may need to multi-task and monitor multiple sessions \cite{Sato2023}, such tools can reduce the cognitive load for interpreting the nonverbal behavior of multiple social scenes concurrently. Another use case is in facilitating interactions with users joining at different levels of immersion (e.g., audio only, full body tracking). For example, model that predict and notify audio-only users of the speaking intentions of immersive users can help them better navigate social scenarios. Conversely, systems can model nonverbal behaviors from audio-only users and notify fully-immersed users of possible speaking intentions of audio-only users, thereby addressing the asymmetry in immersion.

Another application is in training user awareness in detecting turn-taking behaviors. One option is to visually guide users to notice predictive nonverbal features before speaking turns, similar to immersive training systems \cite{Huang2021, Brough2007}. Such training can be beneficial to populations who struggle socially \cite{Freeman2002, Hall1999} and professionals in domains such as education and healthcare. Our features analysis on individual and group differences also offers implications on group composition. Namely, aligning with works on social dynamics and individual differences \cite{Ramsay1968, Beukeboom2013, Aran2013, Lepri2010}, practitioners interested in shaping turn-taking behaviors can vary group size and leverage individuals characteristics to form the ``ideal social group.''

Finally, our work sheds light on virtual agents instantiation. Conversational agents struggle with interrupting at appropriate times \cite{Maier2022CA}, and we foresee future systems leveraging turn-taking predictions to decide whether and when to interject in natural and non-intrusive ways. Possible application of virtual agents include teachers in classrooms, moderators in focus groups and conferences, and facilitators in support groups.

\subsection{Theoretical Implications of Turn-Taking Behaviors in VR}

Understanding how features predict turn-taking behaviors offer practical benefits (e.g., smaller models with curated features) and theoretical implications (\textbf{RQ3}). Here, we expand on the latter, drawing parallels to past literature and supplementing them with post-hoc analysis.

The results on individual and group differences extend works demonstrating that personalities such as extraversion, agreeableness, and neuroticism are correlated with user’s verbal behaviors \cite{Lepri2010, Aran2013}. As noted in Appendix \ref{apx: stats logistic regression}, features related to personalities and group sizes significantly contributed to logistic regression models in at least one of the three tasks. Additionally, we found that more extroverted listeners are predicted to more likely be the next speaker, corroborating past findings that extraversion is related to speaking time \cite{Lepri2010} and speaking turns \cite{Aran2013}. Our analysis on group size suggested that listeners in three-person groups are more likely to be the next speaker compared to listeners in four-person groups. Though we focused on individual feature interpretation, it is also possible that models used group size in combination with motion-related features for their predictions, as group size can be related to head orientations \cite{Hadley2021}. Another related theoretical thread to consider regarding group size is how differences in group composition (e.g., personality, demographics) may affect social dynamics and psychological processes differently. For example, larger groups, compared to smaller ones, tend to report lower levels of group identity and greater role differentiation in performing group tasks, and consequently require more control from leadership to coordinate efforts across group members \cite{Hare1981}. As such, with many of the collaborative activities participants engaged in within our dataset (see Appendix \ref{dataset details} for the list of discussion and group activities), we imagine more drastic and fundamental differences in group behaviors, both verbally and nonverbally, across group that greatly differ in size. While we studied three to four person social groups, just how comparable and similar are turn-taking behaviors between groups examined in this paper and those from much larger groups (e.g., over 10 people)? These findings and considerations highlight the need to consider individual and group characteristics when predicting social dynamics and instantiating virtual agents with distinct personas and natural behaviors. One possible approach is to continue leveraging VR to study large and diverse demographics with high variances in individual differences, and doing so in controlled settings \cite{WangChapterVRTool}.

Speech-related features exhibited high importance and salient effects on probability estimates for the gradient boosting classifiers, and were also significant predictors for the logistic regressions. Furthermore, our models predicted that those who more recently spoke, both in terms of speaking turns and time, are more likely to be the next speaker. While we did not analyze speech sequence features given the likely interactions between turn indices \cite{Parker1988}, it provided important signals to predicting speech behaviors, in particular to next speaker prediction.

Our results revealed that group visual attention behavior is useful in distinguishing between turn transition and continuing speech, with the individual that the group is looking more directly towards predicted to be more likely the upcoming speaker. These findings extend past insights that visual attention measurements are useful in modeling speech behaviors \cite{Jokinen2013, Jokinen2009} and classroom discourse \cite{Stark2024} by demonstrating that the social group’s direct visual attention towards individuals is still predictive of the next speaker even when VR groups are not constrained to fixed virtual positions. While features related to direct visual attention between the listener and previous speaker were significant in predicting the next speaker in logistic regressions, they did not exhibit high importance for gradient boosting models. Nevertheless, these features could still offer insights on speech behaviors when sampled at other moments, for example during speech events or immediately after speaking turns.

Our models predicted that listeners who engaged in substantial rotation in head pitch (e.g., nodding, abruptly looking up or down) and moved their heads and left hands vertically upward exhibited more speaking intentions. Our feature significance results largely echoed the importance of these features. An additional analysis detailed in Appendix \ref{two-variable head pitch analysis} on head pitch revealed that the combination of high instantaneous speeds in both the upward and downward directions can be further indicative of speaking intentions. We also found that the previous speaker’s left hand y-axis position and the head y-axis position and pitch rotation influenced performance, though their feature significance and effects on probability estimates are less consistent across tasks. Of note, features related to the previous speaker's left hand y-axis position were not significant in the logistic regressions. Possible explanations include their potential interactions with other features and non-linear contributions to the predictions. The inconsistency across tasks could be due to the different roles the previous speaker plays in the tasks. For example, while the previous speaker’s egocentric motion could help predict whether they will continue to speak, it is not informative for predicting which listener will speak next. One explanation for the finding on left-hand motion is its association with users raising their hands to access the menu, tablet, and audio button. These findings corroborate past research demonstrating the predictive capabilities of tracking data for modeling speech behavior \cite{Ishii2017Head, Chen2024}, and contribute insights on how egocentric motion is related to the predicted behaviors.

Importantly, though our findings align with prior literature, feature interpretation can differ across virtual context. Similar to how social factors such as group size and personality influenced our model predictions, virtual context such as room size can also impact nonverbal behaviors \cite{Han2023JCMC, Han2024Spatial}. Platform-specific characteristics such as the location of virtual menus could influence body motion and visual attention preceding speech events. Though our dataset varied group size and spatial context, model performance and interpretation may still differ when evaluated on novel social settings.


\section{Limitations and Future Work}

Our work has several limitations. To start, we did not interpret features related to prior speech sequences or incorporated features related to verbal transcripts. Works should examine how virtual speech sequences differ from face-to-face ones and investigate how they predict turn-taking dynamics. Researchers should also extract additional verbal features using transcripts, for example features using Linguistic Inquiry and Word Count \cite{LIWC}. Relatedly, though we did not encode virtual context characteristics, works should sample turn-taking behaviors in drastically-different virtual spaces and curate approaches for parameterizing them. Researchers should also explore whether incorporating gaze-related features enhances model performance since our dataset did not contain them. To balance model performance, complexity, and interpretability, we used summary statistics and standard machine learning models. We imagine works exploring other deep learning architectures \cite{Wu2021GNN,Yu2019LSTM} to capture additional nuances of group dynamics. Another avenue of future work lies in benchmarking the real-time performance of these models and leveraging predictive models to reduce miscommunication \cite{Akselrad2023} and mitigate verbal harassment \cite{Freeman2022, Freeman2023}.

In our dataset, students gathered in fixed groups using the same social VR platform and embodied avatars that lacked nonverbal cues such as facial expressions. To fully evaluate model robustness, researchers should examine different social platforms with varying levels of avatar representation \cite{Smith2018} and immersion \cite{Abdullah2021}, change group membership over time \cite{Salehi2018}, and vary group sizes \cite{Han2023JCMC}. As noted in Section \ref{sec: model selection}, differences in group sizes beyond the scope of our paper could yield drastic differences in social dynamics and lower the applicability of our approach. Different activity types, for example those varying in virtual mobility (e.g., scavenger hunt vs. discussion), could elicit similar concerns. Future research should therefore probe the limits to improving performance through naively scaling up data and sampling across a more diverse set of group interactions (e.g., group size, activity type). It may also be necessary to consider different machine learning architectures for modeling certain group activities, for example training expert sub-models to handle different social scenarios before integrating them into for making predictions \cite{Lin2024}. Doing so will not only suggest technical improvements to the applicability of predicting turn-taking behaviors but also built toward a better understanding of how VR contexts influence social dynamics.

One other limitation pertains to the dataset’s convenience sample of university students. To investigate the generalizability of our insights, researcher should study more representative demographics. Additionally, discrepancies between virtual and physical motion (e.g., physically seated users with standing virtual avatars) warrant scrutiny for their impact on turn-taking behaviors. Finally, while we examined three turn-transition categories, we excluded backchanneling. As backchanneling and overlap speech make up a key component to social dynamics \cite{Schegloff2000} and have unique challenges in CMC due to latencies and diminished nonverbal behaviors \cite{Smith2018, Sheng2021, Seuren2021}, works should investigate overlapping speech, for example through predicting unwanted interruptions. 

\section{Conclusions}

Being able to predict turn taking in VR affords opportunities for understanding immersive social interactions and enables systems to administer support and intervention. In this work, we studied turn-taking behaviors of student engaging in open-ended group activities in VR over four weeks. We predicted turn-taking behaviors using features describing individual and group characteristics and extracted features concerning speech-related behaviors, egocentric motion, and dyadic and group relationships. We found that gradient boosting classifiers achieved the best performance, considerably better than prediction by chance. Additional analysis revealed that listener personality, group size, group visual attention, and listener and previous speaker’s head pitch, head y-axis position, and left hand y-axis position were key features affecting performance and predictions. Our results suggest that these features are reliable indicators, as models were robust when evaluated on unseen activities, weeks, and groups. Taken together, our work contributed a better understanding of how tracking data and individual and group characteristics can predict VR turn taking. We believe our insights will motivate research on modeling social dynamics, and support practitioners to use behavioral predictions to deliver assistance and training in facilitating immersive social interactions.

\begin{acks}
Acknowledgements withheld for the review process.
\end{acks}

\bibliographystyle{ACM-Reference-Format}
\bibliography{sample-base}

\appendix

\newpage
\section{Descriptions of Weekly Group Activities} \label{dataset details}

\begin{table*}[h]
  \small
  \begin{threeparttable}
  \caption{Descriptions of Weekly Group Discussion and Activities.}
  \label{tab: activity description}
  
  \renewcommand{\arraystretch}{1.8}
  \begin{tabular}{cp{0.11\linewidth}p{0.77\linewidth}}
  
    Week & Topic & Discussion Agenda Items \& Prompts \\
    \midrule
    1 & Accessibility &
    \begin{minipage}[t]{\linewidth}
      \begin{itemize}[leftmargin=*, nosep]
        \item Introduce yourself (i.e., name, year, major, what are you most excited about learning in this class, favorite thing you have done in VR)
        \item Talk through the preliminary ideas for your ``Built VR world/scene'' project
        \item Discuss accessibility within the context of ENGAGE (e.g., what are the constraints?)
        \item List things that ENGAGE does well vs. does not do well (e.g., using sticky notes)
      \end{itemize}
    \end{minipage} \\
    2 & Avatars &
    \begin{minipage}[t]{\linewidth}
      \begin{itemize}[leftmargin=*, nosep]
        \item Consider the templates of storyboards we've provided for your storyboard assignment. What are some elements you are considering including in your storyboard? How do you plan on using the affordances unique to VR, such as presence, the ability to move around in 3D space, spatialized sound, etc.? Are you planning on showcasing this in your storyboard?
        \item Reimagine what your avatar would look like. Either draw an avatar that you wish represents you or an avatar you would like to embody. This can, but doesn’t have to, be a human avatar.
        \item Show-and-tell for created designs

      \end{itemize}
    \end{minipage} \\
    3 & Medical & 
    \begin{minipage}[t]{\linewidth}
      \begin{itemize}[leftmargin=*, nosep]
        \item Consider the medical applications we learned in the readings, class, and meditation AltspaceVR journey. What was the most surprising, promising, or concerning? How does this class change your perception of using VR for medical purposes?
        \item Collaboratively work with your group members to create a meditation room or a safe space using any of the ENGAGE tools (e.g., 3D pen, IFX, sticky notes)

      \end{itemize}
    \end{minipage} \\
    4 & Education & 
    \begin{minipage}[t]{\linewidth}
      \begin{itemize}[leftmargin=*, nosep]
        \item Consider a target audience/population (e.g., students of a certain age group, students with a certain learning disability, older students). 
        \item Consider a goal (e.g., retaining factual information, having students experience something) 
        \item Consider a topic of interest (e.g., language, STEM, social skills)
        \item Empathize, define, ideate, and prototype an application tailored to your audience, goal, and topic. Have a member of your group test out/role-play a student using the application

      \end{itemize}
    \end{minipage} \\
    
  \end{tabular}
  \end{threeparttable}
\end{table*}

\newpage

\section{Feature Importance for All Features} \label{apx: full feature importance}

\begin{figure*}[h!]
 \centering 
 \includegraphics[width=0.88\textwidth,keepaspectratio]{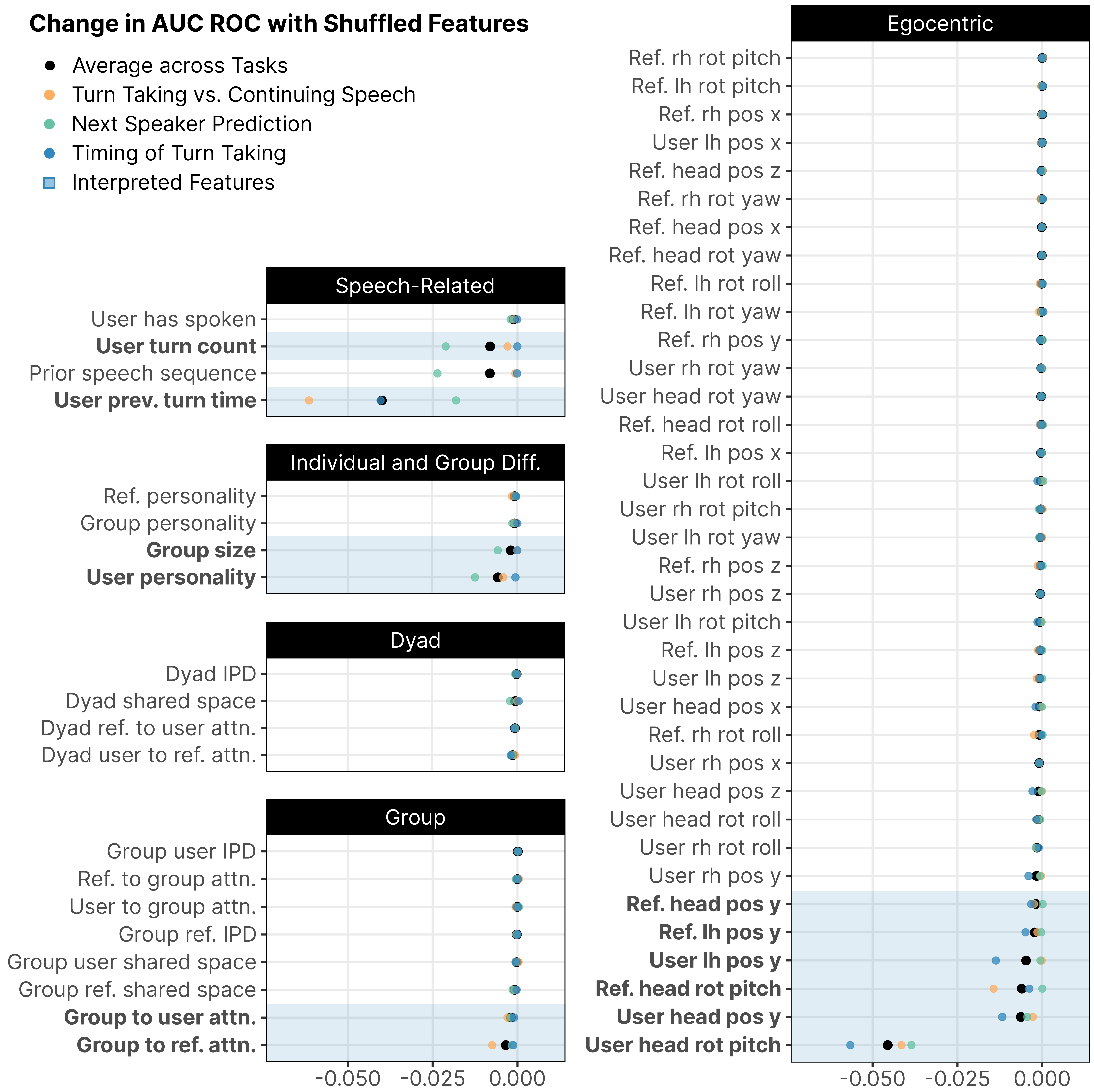}
 \caption{Feature Importance Analysis. Each panel reports the change in AUC of the ROC after shuffling features across the three prediction tasks for a given feature group. Black dots denote the average drop in accuracy across three tasks. Features within each panel are ordered in increasing average drop in accuracy from top to bottom. Bolded features and their corresponding rows with light blue backgrounds indicate features that are further interpreted in Section \ref{sec: feature interpretation}. Diff. = differences; Ref. = reference user; prev. = previous; IPD = interpersonal distance; attn. = visual attention; lh = left hand; rh = right hand.}
 \label{fig: feature importance full}
\end{figure*}

\newpage

\section{Wald Tests for Logistic Regression Models} \label{apx: stats logistic regression}

We present results from joint Wald-tests ran on the trained logistic regression models, specifically the \emph{p} values from the joint hypothesis tests that all features related to a specific construct do not significantly contribute to the predictive model. The orders of the feature groups and constructs follow that of Appendix \ref{apx: full feature importance}. Features with statistical significance, evaluated at $\alpha=.05$, are bolded.

\begin{table*}[h!]
  \small
  \begin{threeparttable}[b]
  \caption{Results from Wald-tests on Speech-Related Features. We report the $p$ values from the joint hypothesis tests. prev. = previous.}
  \label{tab: speech related stats}
  \begin{tabular}{p{0.25\linewidth}ccc}

    & Turn Taking vs. & Next Speaker & Timing of\\
    & Continuing Speech & Prediction & Turn Taking \\
    \midrule
     User has spoken & \textbf{<.001} & \textbf{<.001} & .264 \\
User turn count & \textbf{.046} & \textbf{<.001} & \textbf{.004} \\
Prior speech sequence & \textbf{<.001} & \textbf{<.001} & \textbf{<.001}  \\
User prev. turn time & \textbf{<.001}  & \textbf{.046} & \textbf{<.001}  \\
     
  \end{tabular}
  \end{threeparttable}
\end{table*}

\begin{table*}[h!]
  \small
  \begin{threeparttable}[b]
  \caption{Results from Wald-tests on Individual and Group Differences Features. We report the $p$ values from the joint hypothesis tests. Ref. = reference user.}
  \label{tab: individual and group related stats}
  \begin{tabular}{p{0.25\linewidth}ccc}

    & Turn Taking vs. & Next Speaker & Timing of\\
    & Continuing Speech & Prediction & Turn Taking \\
    \midrule
Ref. personality & \textbf{.020} & \textbf{<.001} & \textbf{<.001} \\
Group personality & .189 & \textbf{<.001} & .191 \\
Group size & \textbf{.001} & \textbf{<.001} & .646 \\
User personality & \textbf{.001} & \textbf{<.001} & .189 \\
     
  \end{tabular}
  \end{threeparttable}
\end{table*}

\begin{table*}[h!]
  \small
  \begin{threeparttable}[b]
  \caption{Results from Wald-tests on Dyad Features. We report the $p$ values from the joint hypothesis tests. Ref. = reference user; IPD = interpersonal distance; attn. = visual attention.}
  \label{tab: dyad related stats}
  \begin{tabular}{p{0.25\linewidth}ccc}

    & Turn Taking vs. & Next Speaker & Timing of\\
    & Continuing Speech & Prediction & Turn Taking \\
    \midrule
Dyad IPD & \textbf{.004} & .066 & .095 \\
Dyad shared space & .177 & .096 & .628 \\
Dyad ref. to user attn. & .846 & \textbf{<.001} & .057 \\
Dyad user to ref. attn. & .246 & \textbf{<.001} & .087 \\
     
  \end{tabular}
  \end{threeparttable}
\end{table*}

\begin{table*}[h!]
  \small
  \begin{threeparttable}[b]
  \caption{Results from Wald-tests on Group Features. We report the $p$ values from the joint hypothesis tests. Ref. = reference user; IPD = interpersonal distance; attn. = visual attention.}
  \label{tab: group stats}
  \begin{tabular}{p{0.25\linewidth}ccc}

    & Turn Taking vs. & Next Speaker & Timing of\\
    & Continuing Speech & Prediction & Turn Taking \\
    \midrule
     
Group user IPD & \textbf{<.001} & .588 & .276 \\
Ref. to group attn. & \textbf{<.001} & \textbf{<.001} & \textbf{.016} \\
User to group attn. & \textbf{<.001} & .298 & \textbf{.002} \\
Group ref. IPD & \textbf{.009} & .347 & .056 \\
Group user shared space & \textbf{.017} & .982 & .332 \\
Group ref. shared space & \textbf{<.001} & .784 & .292 \\
Group to user attn. & \textbf{<.001} & .080 & \textbf{.007} \\
Group to ref. attn. & \textbf{<.001} & \textbf{<.001} & .070 \\
     
  \end{tabular}
  \end{threeparttable}
\end{table*}

\begin{table*}[h]
  \small
  \begin{threeparttable}[b]
  \caption{Results from Wald-tests on Egocentric Features. We report the $p$ values from the joint hypothesis tests. Ref. = reference user; prev. = previous; IPD = interpersonal distance; attn. = visual attention; lh = left hand; rh = right hand.}
  \label{tab: egocentric stats}
  \begin{tabular}{p{0.25\linewidth}ccc}

    & Turn Taking vs. & Next Speaker & Timing of\\
    & Continuing Speech & Prediction & Turn Taking \\
    \midrule
Ref. rh rot pitch & \textbf{.028} & .964 & .297 \\
Ref. lh rot pitch & .195 & .993 & \textbf{.014} \\
Ref. rh pos x & .143 & .997 & \textbf{.007} \\
User lh pos x & \textbf{.015} & .181 & .566 \\
Ref. head pos z & \textbf{.018} & .588 & .505 \\
Ref. rh rot yaw & \textbf{<.001} & .984 & .187 \\
Ref. head pos x & .154 & .541 & .190 \\
Ref. head rot yaw & .202 & .823 & .277 \\
Ref. lh rot roll & .252 & .991 & .766 \\
Ref. lh rot yaw & .470 & .967 & .197 \\
Ref. rh pos y & .493 & .992 & .103 \\
User rh rot yaw & \textbf{.002} & \textbf{.017} & .173 \\
User head rot yaw & \textbf{.001} & .590 & .343 \\
Ref. head rot roll & .058 & .947 & .191 \\
Ref. lh pos x & \textbf{.007} & .994 & \textbf{.027} \\
User lh rot roll & .961 & .316 & \textbf{.043} \\
User rh rot pitch & .564 & .225 & .665 \\
User lh rot yaw & \textbf{.012} & .052 & \textbf{.035} \\
Ref. rh pos z & .646 & .999 & .149 \\
User rh pos z & .080 & .816 & .595 \\
User lh rot pitch & .110 & .161 & .440 \\
Ref. lh pos z & .198 & .977 & .057 \\
User lh pos z & .185 & .246 & \textbf{.021} \\
User head pos x & .111 & .086 & .054 \\
Ref. rh rot roll & \textbf{<.001} & .971 & .090 \\
User rh pos x & \textbf{.048} & .215 & .558 \\
User head pos z & .719 & .052 & .219 \\
User head rot roll & \textbf{.003} & \textbf{<.001} & \textbf{<.001} \\
User rh rot roll & \textbf{.002} & \textbf{<.001} & \textbf{.027} \\
User rh pos y & \textbf{.035} & .082 & .230 \\
Ref. head pos y & \textbf{<.001} & .605 & \textbf{<.001} \\
Ref. lh pos y & .102 & .942 & .213 \\
User lh pos y & \textbf{.002} & \textbf{.027} & .087 \\
Ref. head rot pitch & \textbf{<.001} & .640 & \textbf{.040} \\
User head pos y & \textbf{<.001} & \textbf{.015} & \textbf{<.001} \\
User head rot pitch & \textbf{<.001} & \textbf{<.001} & \textbf{<.001} \\
     
  \end{tabular}
  \end{threeparttable}
\end{table*}

\newpage
\clearpage

\section{Additional Results on Model Predictions} \label{apx: Additional Results on Model Predictions}

In this section, we present additional analyses on prediction performance across sessions, groups, and weeks. To do this, we first randomly split each of the three datasets, one for each task, into a training set with 90\% of the data and a testing set containing the remaining 10\%. Then, we built models on the training data and evaluated them on the testing set in three ways: across sessions, groups, and weeks. Specifically, when evaluating across weeks, we calculated the average AUC and standard error given the trained models' performances across four test sets, each containing the testing data from a specific week. We repeated this procedure for extracting the AUC and standard errors of the models across sessions and groups. Note that the reported average AUC here gives equal weight to each session, group, and week, which differs from the cross-validation evaluations presented in Section \ref{sec: task prediction results}. 



\begin{table*}[h!]
  \small
  \begin{threeparttable}[b]
  \caption{Model Performance (measured as the AUC of the ROC curve) on Predicting Turn-Taking Behavior vs. Continuing Speech. We report the average AUC and standard error on the testing set partitioned based on sessions, groups, and weeks. An AUC of 0.50 means that the model's ability to distinguish between positive and negative samples is no better than random chance. Bolded numbers denote best performance by metric.}
  \label{tab: task 1 appendix performance}
  \begin{tabular}{p{0.26\linewidth}ccc}
    & \multicolumn{3}{c}{Performance Metrics}\\ 
    \cmidrule{2-4}
  
    Prediction Model& By Session& By Group & By Week\\
    \midrule
     
    Logistic Regression & 0.70 (0.02) & 0.71 (0.02) & 0.73 (0.01)\\
    MLP Classifier & 0.71 (0.02) & 0.70 (0.02) & 0.72 (0.02)\\
    Random Forest Classifier & 0.76 (0.01) & 0.77 (0.01) & 0.79 (0.01)\\
    Gradient Boosting Classifier & \textbf{0.80 (0.01)} & \textbf{0.81 (0.01)} & \textbf{0.82 (0.01)}\\
     
  \end{tabular}
  \end{threeparttable}
\end{table*}

\begin{table*}[h!]
  \small
  \begin{threeparttable}[b]
  \caption{Model Performance (measured as the AUC of the ROC curve) on Next Speaker Prediction. We report the average AUC and standard error on the testing set partitioned based on sessions, groups, and weeks. An AUC of 0.50 means that the model's ability to distinguish between positive and negative samples is no better than random chance. Bolded numbers denote best performance by metric.}
  \label{tab: task 2 appendix performance}
  \begin{tabular}{p{0.26\linewidth}ccc}
    & \multicolumn{3}{c}{Performance Metrics}\\ 
    \cmidrule{2-4}
  
    Prediction Model& By Session& By Group & By Week\\
    \midrule
     
    Logistic Regression & 0.72 (0.02) & 0.72 (0.02) & 0.72 (0.01)\\
    MLP Classifier & 0.72 (0.02) & 0.72 (0.02) & 0.72 (0.01)\\
    Random Forest Classifier & 0.75 (0.02) & 0.75 (0.02) & 0.75 (0.00)\\
    Gradient Boosting Classifier & \textbf{0.75 (0.02)} & \textbf{0.76 (0.02)} & \textbf{0.77 (0.00)}\\
     
  \end{tabular}
  \end{threeparttable}
\end{table*}

\begin{table*}[h!]
  \small
  \begin{threeparttable}[b]
  \caption{Model Performance (measured as the AUC of the ROC curve) on the Timing of Turn Taking. We report the average AUC and standard error on the testing set partitioned based on sessions, groups, and weeks. An AUC of 0.50 means that the model's ability to distinguish between positive and negative samples is no better than random chance. Bolded numbers denote best performance by metric.}
  \label{tab: task 3 appendix performance}
  \begin{tabular}{p{0.26\linewidth}ccc}
    & \multicolumn{3}{c}{Performance Metrics}\\ 
    \cmidrule{2-4}
  
    Prediction Model& By Session& By Group & By Week\\
    \midrule
     
    Logistic Regression & 0.61 (0.02) & 0.62 (0.02) & 0.62 (0.01)\\
    MLP Classifier & 0.64 (0.01) & 0.64 (0.02) & 0.64 (0.021\\
    Random Forest Classifier & 0.69 (0.01) & 0.68 (0.02) & 0.69 (0.01)\\
    Gradient Boosting Classifier & \textbf{0.72 (0.02)} & \textbf{0.71 (0.02)} & \textbf{0.73 (0.01)}\\
     
  \end{tabular}
  \end{threeparttable}
\end{table*}

\newpage
\clearpage

\section{Additional Results on Predicting Continuing Speech and Next Speaker} \label{apx: AB model resutls}

While our formulation of turn-taking behavior prediction tasks regarded turn-taking behavior vs. continuing speech and next speaker prediction as two individual tasks, we further present an alternative task that aims to predict speaking intentions more generally. This task holds practical importance as it addresses the broader question of forecasting who will speak next, be it speaker continuation or a switch to a new speaker. Importantly, this task differs from that of turn-taking behavior vs. continuing as the former seeks to distinguish between upcoming speakers and users who will not speak, whereas the latter focuses on distinguishing between two types of upcoming speakers.

Specifically, we defined positive samples as those sampled immediately preceding a continuing speech or clean turn taking speech event. For clean turn-taking behaviors, we chose the main user as the upcoming speaker and the previous speaker as the reference user. For samples corresponding to continuing speech, we chose the upcoming speaker as both the main and reference user. Negative samples were selected also at moments immediately preceding either a clean turn taking or continuing speech event. For the negative samples, we set the reference user as the previous speaker and the main user as one of the users who will not be the upcoming speaker. Across all samples, when the main user and the reference user are chosen to be the same user, we set the dyad-related features to zero after standardization. We down-ampled the training data such that there is an equal number of positive and negative data. Similar to Section \ref{sec: task prediction results}, we report in Table \ref{tab: binary ab performance} the AUC for the four machine learning models across the performance metrics introduced in Section \ref{sec: machine learning models}.

We extended our binary classification models to multiclass prediction. Specifically, we evaluated each instance before a turn-taking or continuing speech event to predict which group member, including the current speaker, would speak next. Unlike our previous individual-based predictions, this formulation focuses on group-level predictions. Using the trained binary models from the previous step, we evaluated each user in the scene (i.e., chosen user as the main user, previous speaker as the reference user) and selected the user with the highest probability of a positive label as the predicted next speaker. Comparing the predictions with the true next speakers, we calculated the testing data accuracy and further reported separately the accuracies of the testing data with three-person groups and four-person groups. When presenting accuracies for cross-validation evaluations, we also report their standard errors across all folds. Tables \ref{tab: multiclass session cv}--\ref{tab: multiclass week cv} summarize our results.

\begin{table*}[h!]
  \small
  \begin{threeparttable}[b]
  \caption{Binary Classification Model Performance (measured as the AUC of the ROC curve) on Predicting Continuing and Speech Next Speaker. For metrics using cross validation, which we denote using the subscript cv, we report the average and standard error across all folds. An AUC of 0.50 means that the model's ability to distinguish between positive and negative samples is no better than random chance. Bolded numbers denote best performance by metric.}
  \label{tab: binary ab performance}
  \begin{tabular}{p{0.26\linewidth}cccc}
    & \multicolumn{4}{c}{Performance Metrics}\\ 
    \cmidrule{2-5}
  
    Prediction Model& Session\textsubscript{cv}& Group\textsubscript{cv} & Week\textsubscript{cv} & Week 4\\
    \midrule
     
    Logistic Regression & 0.74 (0.01) & 0.73 (0.01) & 0.74 (0.01) & 0.73\\
    MLP Classifier & 0.75 (0.00) & 0.74 (0.01) & 0.74 (0.01) & 0.74\\
    Random Forest Classifier & 0.79 (0.00) & 0.79 (0.01) & 0.79 (0.01) & 0.79\\
    Gradient Boosting Classifier & \textbf{0.82 (0.00)} & \textbf{0.81 (0.01)} & \textbf{0.82 (0.00)} & \textbf{0.81}\\
     
  \end{tabular}
  \end{threeparttable}
\end{table*}

\begin{table*}[h!]
  \small
  \begin{threeparttable}[b]
  \caption{Multiclass Classification on Continuing Speech and Next Speaker Prediction using Session-Based Cross-Validation.  We report accuracy across testing samples and, in parenthesis, the standard errors of the accuracies across all folds. Bolded numbers denote best performance by metric.}
  
  \label{tab: multiclass session cv}
  \begin{tabular}{p{0.26\linewidth}ccc}
    & \multicolumn{3}{c}{Performance Metric: Session\textsubscript{cv}}\\ 
    \cmidrule{2-4}
  
    Prediction Model& All & Three-Person & Four-Person\\
    \midrule
     
    Logistic Regression & 0.51 (0.01) & 0.54 (0.01) & 0.48 (0.01)\\
    MLP Classifier & 0.54 (0.01) & 0.57 (0.01) & 0.51 (0.01)\\
    Random Forest Classifier & 0.57 (0.01) & 0.60 (0.01) & 0.54 (0.01)\\
    Gradient Boosting Classifier & \textbf{0.62 (0.01)} & \textbf{0.65 (0.01)} & \textbf{0.59 (0.01)}\\
     
  \end{tabular}
  \end{threeparttable}
\end{table*}

\begin{table*}[h!]
  \small
  \begin{threeparttable}[b]
  \caption{Multiclass Classification on Continuing Speech and Next Speaker Prediction using Group-Based Cross-Validation. We report accuracy across testing samples and, in parenthesis, the standard errors of the accuracies across all folds. Bolded numbers denote best performance by metric.}
  \label{tab: multiclass group cv}
  \begin{tabular}{p{0.26\linewidth}ccc}
    & \multicolumn{3}{c}{Performance Metric: Group\textsubscript{cv}}\\ 
    \cmidrule{2-4}
  
    Prediction Model& All & Three-Person & Four-Person\\
    \midrule
     
    Logistic Regression & 0.51 (0.01) & 0.53 (0.02) & 0.49 (0.02)\\
    MLP Classifier & 0.52 (0.01) & 0.55 (0.01) & 0.50 (0.02)\\
    Random Forest Classifier & 0.57 (0.01) & 0.58 (0.02) & 0.55 (0.02)\\
    Gradient Boosting Classifier & \textbf{0.61 (0.01)} & \textbf{0.63 (0.01)} & \textbf{0.59 (0.01)}\\
     
  \end{tabular}
  \end{threeparttable}
\end{table*}

\begin{table*}[h!]
  \small
  \begin{threeparttable}[b]
  \caption{Multiclass Classification on Continuing Speech and Next Speaker Prediction using Week-Based Cross-Validation.  We report accuracy across testing samples and, in parenthesis, the standard errors of the accuracies across all folds. Bolded numbers denote best performance by metric.}
  \label{tab: multiclass week cv}
  \begin{tabular}{p{0.26\linewidth}ccc}
    & \multicolumn{3}{c}{Performance Metric: Week\textsubscript{cv}}\\ 
    \cmidrule{2-4}
  
    Prediction Model& All & Three-Person & Four-Person\\
    \midrule
     
    Logistic Regression & 0.51 (0.01) & 0.54 (0.01) & 0.48 (0.01)\\
    MLP Classifier & 0.54 (0.01) & 0.57 (0.01) & 0.49 (0.00)\\
    Random Forest Classifier & 0.57 (0.01) & 0.60 (0.01) & 0.54 (0.00)\\
    Gradient Boosting Classifier & \textbf{0.62 (0.01)} & \textbf{0.65 (0.01)} & \textbf{0.58 (0.00)}\\
    
  \end{tabular}
  \end{threeparttable}
\end{table*}

As shown in Table \ref{tab: binary ab performance}, gradient boosting classifiers yielded the highest performance with an AUC of 0.81--0.82 across the four performance metrics. Similar to results presented in Section \ref{sec: task prediction results}, the random forest classifiers yielded the second best performance with an AUC of 0.79, followed by the MLP classifiers and the logistic regressions at 0.74--0.75 AUC and 0.73--0.74 AUC. Tables \ref{tab: multiclass session cv}--\ref{tab: multiclass week cv} revealed similar trends, namely that the gradient boasting classifiers yielded the highest accuracy across all evaluation metrics (i.e., 59--65\%). We also found that the accuracies among three-person groups are higher than that for four-person groups. For reference, random chance predictions would result in an accuracy of 33.33\% for three-person groups and 25.00\% for four-person groups.

\newpage
\clearpage

\section{Two-Variable Partial Dependence Plots for Maximum Upward and Downward Speed for Head Pitch} \label{two-variable head pitch analysis}

\begin{figure}[h!]
 \centering 
 \includegraphics[width=0.84\textwidth,keepaspectratio]{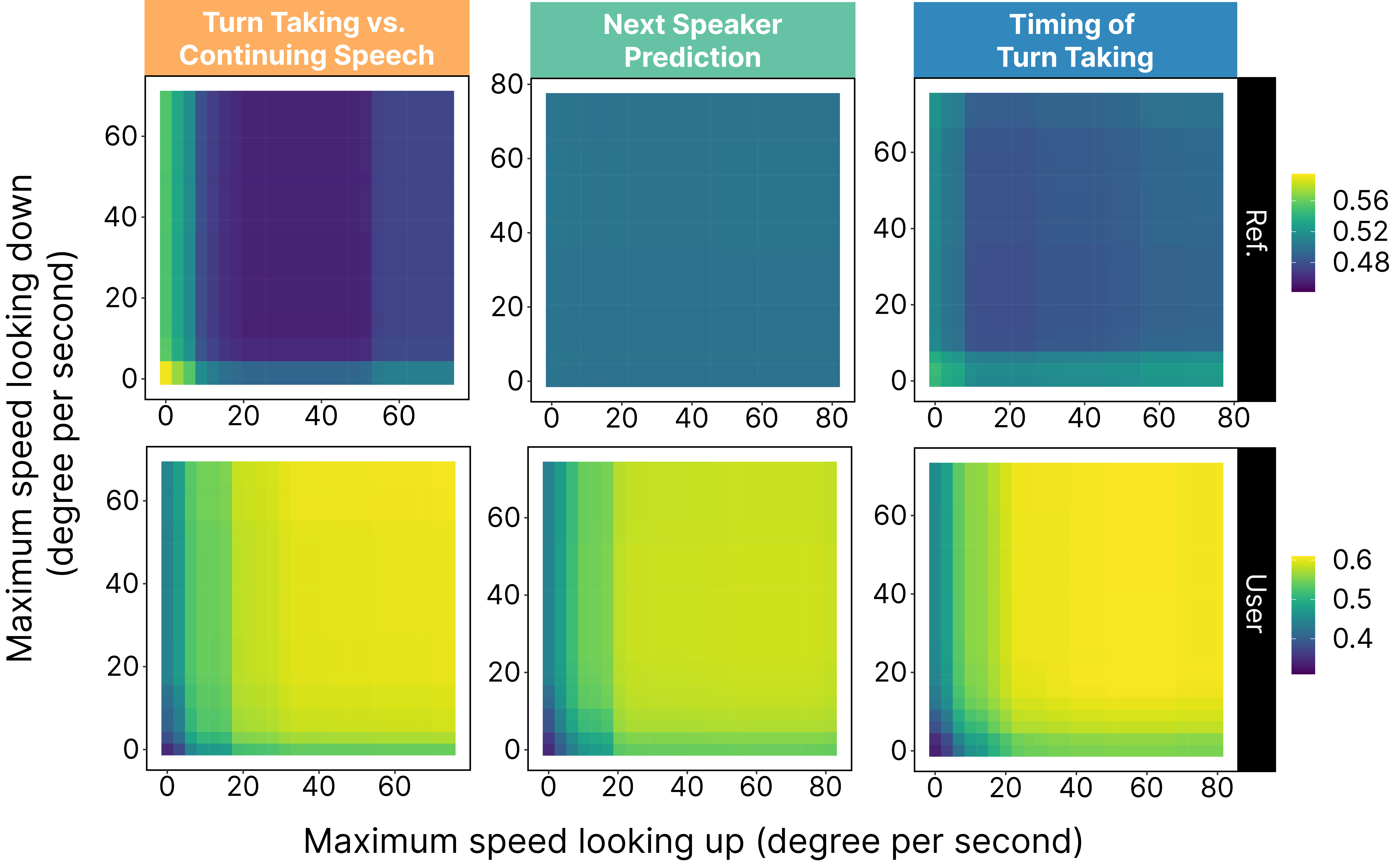}
 \caption{Two-Variable Partial Dependence Plots for Maximum Speed for Head Pitch. The y- and x- axes quantify the maximum speed of the user looking downward and upward, respectively. The y-axis represents values for the maximum head pitch velocity and the x-axis represents values for the minimum head pitch velocity. We negated the original velocity values in the x-axis so that greater values in both axes represent greater instantaneous rotation. The plotted values denote the average probability estimates after setting the two features to their corresponding values based on their coordinates. Ref. = reference user.}
 \label{fig: two-variable partial dependence plot}
\end{figure}

Our results indicated that the extent to which an individual rotates their head in the pitch axis (i.e., velocity upward, velocity downward) is helpful in predicting speaking intentions. To understand how the features for upward head rotation (i.e., minimum head pitch velocity) and downward head rotation (i.e., maximum head pitch velocity) are related, we created two-variable partial dependence plots. Similar to the partial dependence plots we presented in Figures \ref{fig: partial dependencies} and \ref{fig: partial dependencies cont'd}, two-variable partial dependence is derived by first varying the values to the two features and then calculating the average probability estimates of the trained model after this procedure.

Figure \ref{fig: two-variable partial dependence plot} shows the probability estimates after varying the extent of head pitch rotation for the reference and main users across the three tasks. Notably, we see that for all tasks, the combination of greater maximum speed in looking both upward and downward for the main user yielded higher probability of main user speaking intentions. Visually, it also appears that the probability estimates are noticeably higher when the maximum speeds for looking upward and downward are both greater than $\approx$10 degrees per second. This suggests that motions that yield ``high enough’’ values of upward and downward pitch velocities (e.g., nodding) can be indicative of greater speaking intentions. The reverse is true for the tasks of continuing speech vs. turn transition and timing of turn transition regarding the previous speaker, with greater amount of upward and downward head rotation for the previous speaker being associated with smaller probability estimates for the main user's speaking intentions. There was no distinct pattern observed for the previous speaker's head pitch velocities in predicting the next speaker.

\end{document}